\documentclass{aa}  

\usepackage{graphicx}
\usepackage{txfonts}
\usepackage{soul}

\usepackage{hyperref}
\hypersetup{
    colorlinks=true,
    citecolor={blue}
}
\newcommand{\Vband}{{\mathchoice{}{}{\scriptscriptstyle}{} V}}
\newcommand{\Xband}{{\mathchoice{}{}{\scriptscriptstyle}{} X}}
\newcommand{\IPulse}{{\mathchoice{}{}{\scriptscriptstyle}{} I}}
\newcommand{\Pulse}{{\mathchoice{}{}{\scriptscriptstyle}{} P}}
\newcommand{\CX}{{\mathchoice{}{}{\scriptscriptstyle}{} CX}}

\newcommand{\Istk}{{\mathchoice{}{}{\scriptscriptstyle}{} I}}
\newcommand{\Qstk}{{\mathchoice{}{}{\scriptscriptstyle}{} Q}}
\newcommand{\Ustk}{{\mathchoice{}{}{\scriptscriptstyle}{} U}}
\newcommand{\GJ}{{\mathchoice{}{}{\scriptscriptstyle}{} GJ}}

\begin{document}

%\title{Crab Pulsar X-ray Polarimetry: Phenomenological Modeling and Variability Analysis}
\title{Crab Pulsar: IXPE Observations Reveal Unified Polarization Properties Across Optical and Soft X-Ray Bands}
%\subtitle{I. Overviewing the $\kappa$-mechanism}
%%%%%%%%%%%%%%%%%%%%%%%%%%%%%%%%%%%%%%%%%%%%%%%%%%%%%%
%%%%%%%%%%%%%%%%%%%%% OTHER TITLES %%%%%%%%%%%%%%%%%%%
%%%%%%%%%%%%%%%%%%%%%%%%%%%%%%%%%%%%%%%%%%%%%%%%%%%%%%

%Phase-Dependent Polarized Emission of the Crab Pulsar: IXPE Observations Reveal Optical and X-ray Correlations

%IXPE Data Reveal Unified Polarization Patterns in Optical and X-Ray Emission from the Crab Pulsar

%IXPE Observations Reveal Unified Polarization Properties of the Crab Pulsar Across Optical and X-Ray Bands

%Revealing Unified Polarization Properties of the Crab Pulsar Across Optical and X-ray Bands

%Discovery of Unified Polarization Properties Across Optical and X-ray Bands from the Crab Pulsar

%Crab Pulsar: Discovery of Unified Polarization Properties Across Optical and X-ray Bands

%Crab Pulsar: Unifying Polarization Properties Across Optical and X-ray Bands

%Crab Pulsar: IXPE Observations Reveal Unified Polarization Properties Across Optical and X-Ray Bands

\titlerunning{Crab Pulsar X-ray Polarimetry}
\authorrunning{D. Gonz\'alez-Caniulef et al.}

\author{Denis~Gonz\'alez-Caniulef \inst{1}
\and Jeremy~Heyl\inst{2}
\and Sergio~Fabiani\inst{3}
\and Paolo~Soffitta\inst{3}
\and Enrico~Costa\inst{3}
\and Niccol\`o~Bucciantini\inst{4,5,6}
\and Demet Kirmizibayrak\inst{2}
\and Fei Xie\inst{7,3}
}

\institute{
Institut de Recherche en Astrophysique et Plan\'etologie, UPS-OMP, CNRS, CNES, 9 avenue du Colonel Roche, BP 44346 31028, Toulouse CEDEX 4, France\\
\email{dgonzalez-ca@irap.omp.eu}
\and University of British Columbia, Vancouver, BC, Canada
\and INAF Istituto di Astrofisica e Planetologia Spaziali, Via del Fosso del Cavaliere 100, 00133 Roma, Italy
\and INAF Osservatorio Astrofisico di Arcetri, Largo Enrico Fermi 5,
50125 Firenze, Italy
\and Dipartimento di Fisica e Astronomia, Università degli Studi di Firenze, Via Sansone 1, 50019 Sesto Fiorentino (FI), Italy
\and Istituto Nazionale di Fisica Nucleare, Sezione di Firenze, Via Sansone 1, 50019 Sesto Fiorentino (FI), Italy
\and Guangxi Key Laboratory for Relativistic Astrophysics, School of Physical Science and Technology, Guangxi University, Nanning 530004,
China
}

%\date{Received September 15, 1996; accepted March 16, 1997}
\date{Received ... ; accepted ...}

% \abstract{}{}{}{}{} 
% 5 {} token are mandatory
 
\abstract
{We present a phase-dependent analysis of the polarized emission from the Crab pulsar based on three sets of observations by the Imaging X-ray Polarimetry Explorer (IXPE). We found that a phenomenological model involving a simple linear transformation of the Stokes parameters adequately describes the IXPE data.  This model enables us to establish, for the first time, a connection between the polarization properties of the Crab pulsar in the optical and soft X-ray bands, suggesting a common underlying emission mechanism across these bands, likely synchrotron radiation. In particular, the phase-dependent polarization degree in X-rays for pure pulsar emission shows similar features but is reduced by a factor $\approx (0.46-0.56)$ compared to the optical band  (accounting for the contribution of the Knot in the optical), implying an energy-dependent polarized emission. 
In addition, using this model, we study the polarization angle swing in X-rays and identify a potentially variable phase-shift at the interpulse relative to the optical band, alongside a phase-shift marginally consistent with zero persisting at the main pulse.
While the  origin of this variability is unknown and presents a new challenge for theoretical interpretation, our findings suggest that the emission mechanism for the main pulse is likely located far from the neutron star surface, perhaps near or beyond the light cylinder, rather than operating in the inner magnetosphere where vacuum birefringence is expected to be at work. Ignoring the phase-shifts would result in identical phase-dependent polarization angles between the optical and X-ray bands for pure pulsar emission.}


   \maketitle
%
%-------------------------------------------------------------------

\section{Introduction}

The recently launched (December 2021) Imaging X-ray Polarimetry Explorer mission (IXPE) opened a new window to study polarized, soft X-ray sources (2--8 keV) in extreme astrophysical environments \citep{Weisskopf22}. One of the main targets of IXPE Long-term plan (LTP) is the Crab pulsar wind nebula (PWN), one of the most studied  X-ray sources that has also been subject to polarimetric analysis with various instruments \citep[e.g.,][]{weisskopf78, Forot2008, Chauvin2013, Chauvin2017, Vadawale2018, Feng2020, Long2021, Li2022}.  It consists of a fast spinning ($P=33.7$~ms) and highly magnetized ($B=3.8\times10^{12}$~ G) neutron star (Crab pulsar, PSR B0531+21, PSR J0534+2200) that accelerates particles that power a PWN (G184.6-5.8). Although they are considered sources of nonthermal radiation: synchrotron, curvature radiation, and inverse Compton processes, the mechanism of pulsar emission and its location remain subject of debate \citep[for reviews see e.g.,][]{Buhler2014,Harding2019,Philippov2022}. As part of the LTP, three sets of observations were performed for the Crab PWN. The first observation, made during cycle 1, has an exposure time of  $\sim 90$~ks, while the second and third observations, requested as follow-up during cycle 2,  have exposure times of $\sim150$~ks and $\sim60$~ks, respectively.  These observations were made on 21 March 2022, 1 April 2023, and 9 October 2023.

Based on the first observation, the IXPE team reported a polarimetric analysis of Crab PWN focusing on the nebula emission, finding an integrated polarization degree $\mathrm{PD}\sim20\%$. 
In addition, preliminary analyses found a low polarized emission from Crab pulsar with maximum $\mathrm{PD}\sim15\%$ at the core of the main peak \citep{Bucciantini2023}, as well as hints of a fast polarization angle swing $\sim 100^\circ - 150^\circ$ \citep{Wong2023}. 
Later, an analysis that included all three IXPE observations of Crab PWN was performed by \cite{Wong2024}, finding consistent results with those reported in \cite{Bucciantini2023}. Additionally, \cite{Wong2024} identified: i) a well defined S-shaped polarization angle swing at the main peak, ii) polarization in six bins in the main pulse and two phase bins in the interpulse (out of a total of 20 phase bins), and iii) substantial differences in the phase-dependent polarization properties for both the main peak and interpulse compared to those observed in the optical band \citep[][]{slowikowska09}. In particular, they conclude that different emission mechanisms or locations may explain the optical and X-ray polarized emission. 
In contrast, as discussed later, our analysis suggests a common emission mechanism across both bands, differing from the conclusions of \cite{Wong2024} and offering new insights into the Crab pulsar's magnetospheric emission.

Our main goal is to examine the pulsed emission from the Crab pulsar, with a particular focus in the study of the polarization angle swing observed by IXPE. The high magnetic field of Crab pulsar, 
%\st{$B=3.8\times10^{12}$~ G,} 
short rotational period, %\st{$P=33.7$~ms}
%\citep{Hester2008,Buhler2014},
relatively close distance of $\sim2$~kpc \citep{Trimble1973}, and bright pulsed X-ray flux $F{(2-10\,\mathrm{keV})}\sim2.7\times10^{-9}~\mathrm{erg\,cm^{-2}\,s^{-1}}$, make it an ideal X-ray source to study quantum electrodynamics (QED) effects. A strong magnetic field is expected to modify the properties of the vacuum by inducing the temporary formation of virtual electron-positron pairs, which can lead to the appearance of vacuum birefringence \citep{heisenberg36,weisskopf36,Schwinger1951}, a phenomenon that remains experimentally unobserved.  (An astrophysical signature of this phenomenon, however, has been potentially obtained in the optical band for RX J1856.5$-$3754 by \citealt{mignani17}.)  Under vacuum birefringence, electromagnetic waves with different energies and polarization modes propagate at different speeds and decouple at different locations within the pulsar's magnetosphere.  Consequently, if the emission mechanism takes place well inside the light cylinder, it is expected that  rapid rotation of the magnetosphere ($P \lesssim 0.1$\,s) will induce a phase-shift in the polarization angle between observations at different energy bands \citep{heyl00}.

In a fast rotating pulsar, the effects of vacuum birefringence are in competition with the presence of a magnetospheric plasma.  
Assuming a pulsar with a  Goldreich-Julian charge density $n_\mathrm{\GJ}$ \citep{Goldreich1969}, the birefringence effects arising from the vacuum become dominant over those resulting from the plasma density for photon energies that are sufficiently high \citep{heyl00}: 
\begin{equation}
    E > 0.1\,\mathrm{eV} \left( \frac{B}{3.8 \times 10^{12}\,\mathrm{G}} \frac{P}{33.7\,\mathrm{ms}}  \frac{n_\mathrm{\GJ}}{n_e} \right)^{-1/2}
\end{equation}
Therefore, performing observations in optical and higher energy ranges can enable us to investigate vacuum birefringence in magnetized neutron stars, such as  Crab pulsar, as well as  to probe the magnetosphere of these kind of sources.

Extensive studies on the polarization properties of the Crab pulsar have been conducted over the last decades in the optical band \citep[see e.g.][]{cocke1969,Kristian1970,Smith1988,slowikowska09,moran13}. 
(For general reviews on multiwavelength polarimetry of pulsars see also \citealt{mignani2018,Harding2019,Bucciantini2024}.)
The most precise phase-dependent measurements were carried out by \cite{slowikowska09} using OPTIMA at approximately $2$~eV observations. On the other hand, with the IXPE mission operating at sufficiently high-energies, $2-8$~keV range, and sufficiently good sensitivity to carry out phase-dependent polarimetric observations,  it is now possible to search for phase-shifts in the polarization angle of Crab pulsar between optical and X-rays bands. 

In this study, we develop a phenomenological model based on the phase-dependent polarimetric observations of Crab pulsar in the optical band, the closest energy range to soft X-rays where high-quality polarization measurements have been obtained. We find that by applying a linear transformation of the Stokes parameters in the optical band, we can provide a fairly good description of the IXPE phase-dependent observations of the Crab pulsar in the X-ray band. 
This suggests that the underlying emission mechanisms operating across these bands are likely the same. Furthermore,  motivated by the search for a signal of vacuum birefringence \citep{heyl00}, we include two extra free parameters in the model: a phase-shift in the polarization angle swing at both the main pulse and interpulse, and fit them to IXPE observations\footnote{These phase-shifts may also be investigated, for example, in the phase bins with polarization detection reported in \cite{Wong2024}. However, as shown later,  our fit uses all phase bins at a time, making it more sensitive to detect phase-shifts, provided that the model is acceptable.}. We find that, while the phase-shift at the main pulse is marginally consistent with zero for the three IXPE observations of the Crab pulsar, the phase-shift at the interpulse varies between different observations.   
While this variability is likely a new feature in the polarization properties of Crab pulsar in the X-ray, the polarization angle swing in the main peak suggests that the emission mechanism producing the main peak is not affected by vacuum birefringence (within the polarization limiting radius\footnote{ For a dipolar magnetic field, the polarization limiting radius is defined as $r_\mathrm{PL} \approx 1.2\times10^7 \mathrm{cm} \left( \frac{\mu}{10^{30}\,\mathrm{G\,cm}^3}\right)^{2/5}\left( \frac{\nu}{10^{17}\,\mathrm{Hz}}\right)^{1/5} (\sin\theta)^{2/5}$, where 
$\mu$ is the magnetic dipole moment of NS, $\nu$ is the frequency of the radiation,  and $\theta$ is
the angle between the magnetic axis and the line of sight \citep{heyl02}.}) and, therefore, its location is well outside the magnetosphere, perhaps close to or beyond the light cylinder.

The paper is organized as follows. In \S2, we present  our model for the Stokes parameters,  the implementation of the phase-shift in the polarization angle, and the fit to the IXPE data.   Discussion and conclusions are presented in \S3.

\begin{figure*}
\centering
\includegraphics[width=\linewidth]{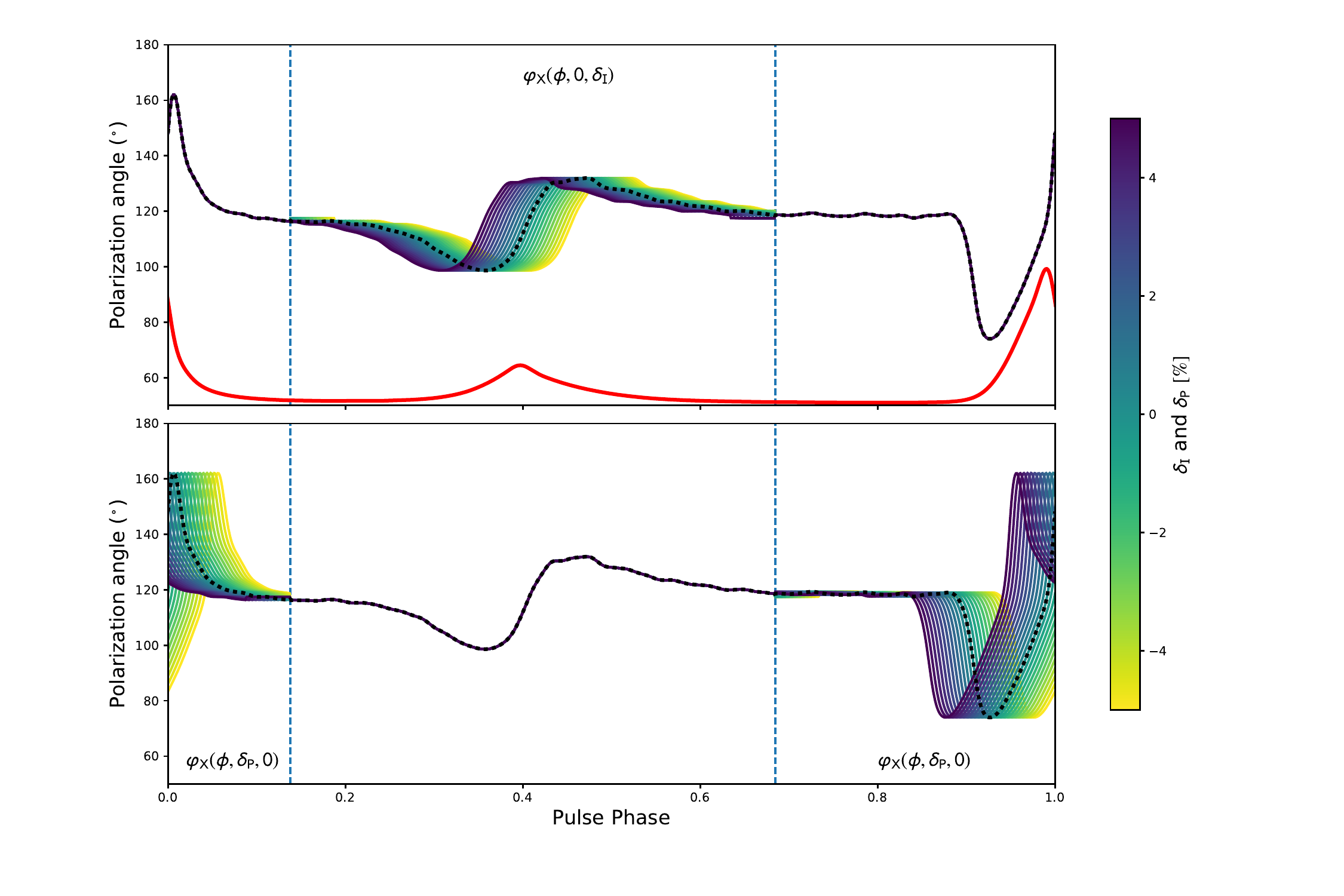}
\caption{Phenomenological polarization angle model, $\varphi_\Xband$, for the pulsar in X-rays considering positive and negative phase-shifts from $-5\%$ to $5\%$ at both the interpulse (upper panel) and main pulse (lower panel). For reference, the black dotted line shows the optical (V-band) polarization  angle  and  the red solid curve  shows the $I$ Stokes  (V-band,  with arbitrary normalization) from \citet{slowikowska09}. Phase shifts are applied within the boundaries indicated by the vertical blue dashed lines.
}
\label{fig:shift_model}
\end{figure*}

\section{Method}
The Crab pulsar is characterized by a double pulsed emission per cycle, which is observable almost across the whole electromagnetic spectrum, from radio waves to high-energy gamma rays \citep[for a review see e.g.][]{Buhler2014}. The locations of the peak for the main pulse and interpulse\footnote{The interpulse is also energy dependent with the ratio of intensity main/interpulse lower in X-ray compared to optical.} show relatively small variations in phase between the optical \citep{slowikowska09} and the X-ray band \citep{weisskopf11}, with a peak-to-peak separation of $\sim0.4$ cycles. In the optical band and higher energies, a {\it bridge}  emission is observed between the peaks. Furthermore, an off-pulse emission (or DC region)  is also typically characterized as that of the pulse profile minimum, where the emission from the Knot (in optical) or PWN (in X-ray) becomes prominent. In order to perform the polarimetric analysis,  previous studies of the Crab pulsar involved subtracting the Stokes parameters associated with the off-pulsed emission from those of the pulsar \citep[e.g.,][]{slowikowska09,Bucciantini2023}. However, we take the opposite approach. We directly extract the pulsed Stokes parameters from the IXPE data, and then we compare them with a pulsar model that accounts for the contribution from the PWN (or any form of constant polarization) as follows.

\subsection{Linear transformation}
In order to perform an in-depth study of the phase-dependent polarization properties of Crab pulsar,  we build a phenomenological model assuming that the main features of the Stokes parameters for the pulsar emission in the optical band \citep{slowikowska09} are preserved in the X-ray band. We also allow the IXPE polarimetric features to stretch or contract when compared to optical observations, which can be expressed as a linear transformation of the Stokes parameters given by
 \begin{equation}
\mathbf{S}_\mathrm{\Xband} \equiv \mathbf{A} \cdot \mathbf{S}_\mathrm{\Vband}  + \mathbf{B},     
 \end{equation}
with 
\begin{equation}
     \mathbf{S}_\mathrm{\Xband}=\begin{pmatrix}  I_\mathrm{\Xband}\\ Q_\mathrm{\Xband} \\ U_\mathrm{\Xband}  \end{pmatrix},\;
     \mathbf{A} =\begin{pmatrix} \alpha & 0 & 0\\0 & \beta & 0\\0 & 0 & \beta \end{pmatrix},\; 
     \mathbf{S}_\mathrm{\Vband}=\begin{pmatrix} I_\mathrm{\Vband}\\ Q_\mathrm{\Vband} \\ U_\mathrm{\Vband}  \end{pmatrix},\;
     \mathbf{B} =\begin{pmatrix} b_\Istk \\ b_\Qstk \\ b_\Ustk\end{pmatrix},
\end{equation}
where the diagonal matrix $\mathbf{A}$ quantifies a deviation from the pulsar's Stokes parameters between optical and X-rays,  while the vector $\mathbf{B}$ quantifies the contribution from the nebula emission or any additional form of constant polarization e.g., spurious polarization (although this should be already removed), and background both celestial and instrumental. For the $\mathbf{A}$ matrix, we repeat the $\beta$ coefficient in the diagonal and we do not consider off-diagonal terms in order to avoid introducing a rotation or mixing of the Stokes parameters of the pulsar between optical and X-ray, which a priori is not justified.   Here, $(I_\mathrm{\Vband},Q_\mathrm{\Vband}, U_\mathrm{\Vband} )$ is the set of Stokes parameters in the optical band (see Appendix \ref{appendix:optical}), which are likely dominated by pulsar emission at almost all phases (with the exception of a DC component as discussed later),  and $( I_\mathrm{\Xband},Q_\mathrm{\Xband}, U_\mathrm{\Xband})$ correspond to the Stokes parameters in the IXPE X-ray band,  which have a relatively large nebula component due to the broad instrumental PSF.

Nevertheless, the $I$ Stokes parameter for the pulsar itself in the X-ray range has already been determined, with minimal contamination from the nebula, through previous observations conducted by Chandra X-ray observatory \citep{weisskopf11}. Therefore, as we want to perform the analysis in the X-rays, we naturally use Crab pulsar's intensity Stokes from Chandra observations (which is different compared to the intensity Stokes from IXPE observations), but preserving the polarimetric characteristics observed in the optical band, i.e., phase-dependent polarization degree and polarization angle. In the linear transformation above, we accomplish this by substituting $\mathbf{S}_\mathrm{\Vband}$ by:
\begin{equation}
     \mathbf{S}^\prime_\mathrm{\Vband}
     =
     \frac{I_\mathrm{\CX}}{I_\mathrm{\Vband}}\mathbf{S}_\mathrm{\Vband}
     =
      p_{\mathrm{\Vband}}I_\mathrm{\CX} \begin{pmatrix} 1/p_{\mathrm{\Vband}} \\ \cos(2\, \psi_\mathrm{\Vband})  \\   \sin(2\, \psi_\mathrm{\Vband})  \end{pmatrix}
\end{equation}
where $I_\mathrm{\CX}$  corresponds to Chandra X-ray intensity of the pulsar (see Appendix \ref{appendix:xray}), and $p_{\mathrm{\Vband}}=\sqrt{ Q_\mathrm{\Vband}^2 + U_\mathrm{\Vband}^2}/I_\mathrm{\Vband}$
and
$\psi_\mathrm{\Vband} = (1/2)\arctan{(U_\mathrm{\Vband}/Q_\mathrm{\Vband})}$  are the the polarization fraction and polarization angle in the optical band, respectively. As discussed in \cite{slowikowska09}, the optical Stokes parameters include a DC component, which is probably due to the PWN knot. However, it is not clear the magnitude of this contribution. Therefore,  we also tested the scenario in which we remove $90\%$ of the DC component from the optical Stokes as follows: 
%\citep[to mimick Figure 5, right panel of][]{slowikowska09}:
\begin{eqnarray}
I_\mathrm{\Vband} &\rightarrow& I_\mathrm{\Vband} - 0.9 I_{\mathrm{\Vband},\mathrm{dc}} \\
Q_\mathrm{\Vband} &\rightarrow& Q_\mathrm{\Vband} - 0.9 Q_{\mathrm{\Vband},\mathrm{dc}} \\
U_\mathrm{\Vband} &\rightarrow& U_\mathrm{\Vband} - 0.9 U_{\mathrm{\Vband},\mathrm{dc}}
\end{eqnarray}
where  $I_{\mathrm{\Vband},\mathrm{dc}}$, $Q_{\mathrm{\Vband},\mathrm{dc}}$, and $U_{\mathrm{\Vband},\mathrm{dc}}$ are the optical Stokes parameters at phase $\approx 0.8$. 

\begin{table*}
\centering
\caption{Summary of the model parameters fitted to three separate observations of the Crab pulsar by IXPE. }
\label{tab:table1}
\begin{tabular}{lcccccccr}
    \hline
    \hline
    \noalign{\vskip 1mm}  
    Obs. & $\alpha/\Delta t$ & $\beta/\alpha$ &  $b_\Istk/\Delta t$ &  $b_\Qstk/\Delta t$ & $b_\Ustk/\Delta t$& $\delta_\mathrm{\IPulse}$ & $\delta_\mathrm{\Pulse}$ & $\chi^2\,/\,\mathrm{dof^{(a)}}$\\    &           &             &  [$\times10^{-2}$ cnt/s] &  [$\times10^{-2}$ cnt/s] &  [$\times10^{-2}$ cnt/s] & $[\%]$ & $[\%]$ & \\    \hline
    \noalign{\vskip 1mm}  

    $1^{\mathrm{st}}$ 
    & $0.930\pm0.004$
    & $ 0.55^{+0.12}_{-0.13}$
    & $72.00\pm0.09$
    & $ 0.62^{+0.42}_{-0.41}$
    & $-16.54^{+0.41}_{-0.41}$
    & $7.42^{+5.94}_{-4.42}$ 
    & $-0.67^{+0.51}_{-0.63}$ & 126.46\,/\,75\\
    \noalign{\vskip 1mm}  
 
    $2^{\mathrm{nd}}$ 
    & $0.957\pm0.003$
    & $ 0.53^{+0.10}_{-0.10}$
    & $76.65\pm0.08$
    & $-0.55^{+0.32}_{-0.32}$
    & $-17.40^{+0.34}_{-0.34}$
    & $21.23^{+3.52}_{-2.74}$ 
    & $-0.90^{+0.57}_{-0.70}$ & 95.65\,/\,75\\
    \noalign{\vskip 1mm}   

    $3^{\mathrm{rd}}$ 
    & $0.913\pm0.007$
    & $ 0.31^{+0.19}_{-0.21}$
    & $76.93\pm0.13$
    & $ 0.24^{+0.56}_{-0.56}$
    & $-16.98^{+0.54}_{-0.55}$
    & $-2.16^{+9.13}_{-13.88}$
    & $-1.13^{+1.61}_{-2.24}$ & 100.75\,/\,75\\
    \noalign{\vskip 1mm}  
    \hline
    \noalign{\vskip 1mm}  
    \multicolumn{9}{c}{Model without phase shift}\\
    \noalign{\vskip 1mm}      
    \hline
    \noalign{\vskip 1mm}  
    $1^{\mathrm{st}}$  
    & $0.930\pm0.004$
    & $ 0.51^{+0.12}_{-0.12}$
    & $72.00\pm0.09$
    & $ 0.68^{+0.41}_{-0.41}$
    & $-16.66^{+0.40}_{-0.42}$
    & 0 & 0 & 132.05\,/\,77\\
    \noalign{\vskip 1mm}

    $2^{\mathrm{nd}}$  
    & $0.957\pm0.003$
    & $ 0.42^{+0.10}_{-0.10}$
    & $76.65\pm0.08$
    & $-0.48^{+0.33}_{-0.34}$
    & $-17.70^{+0.35}_{-0.33}$
    & 0 & 0 & 96.71\,/\,77\\
    \noalign{\vskip 1mm}

    $3^{\mathrm{rd}}$  
    & $0.913\pm0.007$
    & $ 0.44^{+0.16}_{-0.16}$
    & $76.93\pm0.13$
    & $ 0.49^{+0.55}_{-0.52}$
    & $-16.82^{+0.52}_{-0.52}$
    & 0 & 0 & 101.07\,/\,77\\
    \noalign{\vskip 1mm}  
    \hline
    \noalign{\vskip 1mm}  
    \multicolumn{9}{c}{Model including $90\%$ subtraction of DC component from the Stokes parameters in the optical band }\\
    \noalign{\vskip 1mm}      
    \hline
    \noalign{\vskip 1mm}

    $1^{\mathrm{st}}$  
    & $0.930\pm0.004$
    & $ 0.61^{+0.14}_{-0.14}$
    & $72.00\pm0.09$
    & $ 0.47^{+0.40}_{-0.41}$
    & $-17.11^{+0.37}_{-0.37}$
    & $2.91^{+1.36}_{-2.22}$ 
    & $-0.54^{+0.45}_{-0.56}$
    & 123.59\,/\,75\\
    \noalign{\vskip 1mm} 
    
    $2^{\mathrm{nd}}$  
    & $0.957\pm0.003$
    & $ 0.53^{+0.10}_{-0.10}$
    & $76.65\pm0.08$
    & $-0.98^{+0.29}_{-0.31}$
    & $-17.93^{+0.30}_{-0.31}$
    & $32.91^{+1.23}_{-1.51}$ 
    & $-0.99^{+0.57}_{-0.65}$ & 84.33\,/\,75\\
    \noalign{\vskip 1mm}  

    $3^{\mathrm{rd}}$   
    & $0.913\pm0.007$
    & $ 0.54^{+0.18}_{-0.20}$
    & $76.93\pm0.13$
    & $ 0.42^{+0.53}_{-0.54}$
    & $-17.23^{+0.47}_{-0.48}$
    & $-4.51^{+4.41}_{-7.17}$ 
    & $-0.78^{+0.85}_{-1.09}$ & 95.96\,/\,75\\
    \noalign{\vskip 1mm}  
    \hline
\end{tabular}
\raggedright 

$^{(a)}$ $\chi^2$ statistic reported for the set of parameters $\{\beta$, $b_\Qstk$, $b_\Ustk$, $\delta_\mathrm{\IPulse}$, $\delta_\mathrm{\Pulse}$\}. The parameters \{$\alpha$, $b_\Istk$\} are fitted separately to the IXPE Stokes $I$; they do not depend on the treatment of the polarized components of the Stokes vector.  The quantity $\Delta t$ is the exposure time divided by the number of phase bins (40). 
\end{table*}

\subsection{Polarization angle including phase shifts}

If the emission mechanism takes place well inside the light cylinder (or near the NS surface), vacuum birefringence is expected to induce a phase-shift in the polarization angle that depends on both the energy and rotational phase of the pulsar (e.g., with the polarization angle swing for low-frequency emission lagging behind that for higher-frequency radiation).  Phase shifts have been computed, for example, for both a corotating dipole model and a Deutsch model by \cite{heyl00}. Both models show that the phase variation for the polarization angle within the energy range of IXPE is very small compared to the variation between optical V-band and soft X-rays. Therefore, we investigate potential phase shifts for the polarization angle as the difference in phase associated to two representative energies: 
\begin{itemize}
    \item[i)] $E=2$~eV, which corresponds to the optical V-band of Crab pulsar's polarimetric observations \citep{slowikowska09}, and
    \item[ii)] $E=3$~keV, which corresponds   to the energy for the maximum sensitivity of IXPE.
\end{itemize}
As mentioned above, the phase-shift also depends on the rotational phase of the pulsar. In principle, the signal of the phase-shift is stronger at the peak of both the pulse and interpulse, where the polarization angle  also exhibits its maximum swing.
In the following, we define these two peaks to be located within two broad rotational phase intervals: 
\begin{itemize}
    \item[i)]for phases inside the interval $[\phi_1^{\mathrm{min}},\phi_2^{\mathrm{min}}]$ or inter-pulse,   and
    \item[ii)]  for phases outside $[\phi_1^{\mathrm{min}},\phi_2^{\mathrm{min}}]$ or main pulse. 
\end{itemize}
where $\phi_1^{\mathrm{min}}=0.138$ and $\phi_2^{\mathrm{min}}=0.683$ correspond to the two local minimum in intensity Stokes parameter $I_\CX$ as observed by Chandra.  In the case of IXPE observations, around $\phi_1^{\mathrm{min}}$ and $\phi_2^{\mathrm{min}}$  the X-ray emission is substantially affected by the nebula emission. 
Therefore, and for simplicity, we ignore potential continuous phase-dependent phase-shifts. 
Then, for the pulsar in the X-rays,  we implement the phase-dependent polarization angle model including two independent phase-shifts respect to the optical band as follows:
\begin{equation}
  \varphi_\mathrm{\Xband}(\phi,\delta_\mathrm{\Pulse},\delta_\mathrm{\IPulse} )=\begin{cases}
    \psi_\mathrm{\Vband}(\phi + \delta_\mathrm{\IPulse}), & \phi_1^\mathrm{min} < \phi < \phi_2^\text{min},\\
    \psi_\mathrm{\Vband}(\phi + \delta_\mathrm{\Pulse}), &  \text{otherwise},
  \end{cases}
  \label{eq:phase_shifts}
\end{equation}
where $\psi_\mathrm{\Vband}$ corresponds to the phase-dependent polarization angle model from optical observations (V-band), $\phi$ is the rotational phase of the pulsar, and $\delta_\mathrm{\IPulse}$ and $\delta_\mathrm{\Pulse}$ correspond to the phase-shifts at the interpulse and main pulse, respectively. (In the following, positive phase-shifts mean that the polarization angle swing for the optical band is lagging behind the X-ray band.)  From \cite{heyl00}, theoretical expectations for the phase shift due to vacuum birefringence are typically $\delta_\mathrm{\Pulse}= 0.91\%$ and $\delta_\mathrm{\IPulse}= 1.74\%$ at the pulse and interpulse peaks, respectively. Figure \ref{fig:shift_model} shows the model for the polarization angle including negative and positive phase-shifts in the range $[-5,5]\%$.

By including  the phase shifts in the polarization angle, the full optical-to-X-ray linear transformation of the Stokes parameters can be expanded as
\begin{eqnarray}
 I_\mathrm{\Xband} &=&  \alpha\, I_\mathrm{\CX} + b_\Istk \label{eqn:Ix} \\
Q_\mathrm{\Xband} &=& \beta\, p_\mathrm{\Vband}\, 
\cos{\left(2\,\varphi_\mathrm{\Xband}(\delta_\mathrm{\IPulse},\delta_\mathrm{\Pulse})\right)}\,  I_\mathrm{\CX} + b_\Qstk  \label{eqn:Qx}\\
 U_\mathrm{\Xband} &=& \beta\, p_\mathrm{\Vband}\, \sin{\left(2\,\varphi_\mathrm{\Xband}(\delta_\mathrm{\IPulse},\delta_\mathrm{\Pulse})\right)}\,   I_\mathrm{\CX} + b_\Ustk  \label{eqn:Ux}
\end{eqnarray}
Here, the whole set of parameters $\left\{\alpha, \beta,  b_\Istk,  b_\Qstk, b_\Ustk, \delta_\mathrm{\IPulse}, \delta_\mathrm{\Pulse} \right\}$\footnote{The $\alpha$ parameter also absorb missing instrumental effects in the transformation of the unfolded/unabsorved intensity from Chandra to IXPE folded intensity, which explains why  the values of $\alpha/\Delta t$ listed in Table\,\ref{tab:table1} are not exactly 1. 
}
is derived after fitting the model to IXPE polarimetry data. Consequently, the full polarization angle model for the pulsar including nebula emission is  $\psi=(1/2)\arctan{(U_\mathrm{\Xband}/Q_\mathrm{\Xband})}$ and  polarization degree model is
$p=\sqrt{Q_\mathrm{\Xband}^2 + U_\mathrm{\Xband}^2}/I_\mathrm{\Xband}$.

\subsection{Data reduction, fitting procedure, and MCMC}
The data reduction  for the first set of observations is summarized in the method section of \cite{Bucciantini2023}. 
The second and third sets of IXPE observations are reduced in a similar manner, i.e., we use  \textsc{ixpeobssim}  package \citep{Baldini2022} to perform energy calibration, detector WCS correction, and aspect-solution corrections, as well as unweighted analysis. We use a circular subtraction region  of radius 20'' and select photons in the $2-4$ keV range (corresponding to the maximum IXPE sensitivity and to avoid nebula contamination that becomes dominant at higher energies). 
Barycenter correction is obtained with \textsc{barycorr ftool} and photons are phase-folded using a Lomb-Scargle periodogram. 
For the Crab pulsar it is well known that the locations of the main peak at different energy bands are slightly misaligned (of the order of $1\%$ in phase),   with the optical, X-ray, and gamma ray peaks leading the radio peak. In the following, for the main peak we set phase zero at 0.99 relative to the radio peak\footnote{Notice that the radio light curve is significantly different compared to the ones in optical/X-ray.} \citep[see also,][]{Bucciantini2023}.

At variance of \cite{Bucciantini2023}, the binned data are analysed using equi-populated binning to produce a constant MDP$_{99}$ through all rotational phases of the pulsar\footnote{$\mathrm{MDP}_{99} = 4.29/(\mu\sqrt{N})$ is the minimum detectable polarization at $99\%$ confidence level, where $\mu$ is the modulation factor of the detector and $N$ is the number of counts (when background is negligible).}. 
This approach is employed to prevent any modulation of the MDP$_{99}$ that occurs, for example, when using equi-spaced binning in phase (as  Crab pulsar is strongly double-peaked in counts over a rotational cycle). We use the \textsc{pcube} algorithm to produce 40 phase bins for the $I, Q, U$ Stokes parameters as well as polarization degree and angle.  As shown in the second panel of Figure~\ref{fig:binned_Obs1}, all polarization degree data points are above the MDP$_{99}$.

We fit the model presented in the previous section to the IXPE data.  Since we are interested in the polarization properties of the Crab pulsar, minimizing the effects of the modulation contained in the intensity Stokes, we compute a $\chi^2$ statistic as 

\begin{equation}
\chi^2 = \sum\left(\frac{q - q_\mathrm{m}}{\sigma_q}\right)^2 
           + \left(\frac{u - u_\mathrm{m}}{\sigma_u}\right)^2,
\end{equation}
where
\begin{equation}
\label{eq:model_nstokes}
q_\mathrm{m} = Q_\mathrm{\Xband}/I_\mathrm{\Xband}, \qquad \mathrm{and} \qquad u_\mathrm{m} = U_\mathrm{\Xband}/I_\mathrm{\Xband}    
\end{equation}
correspond to the normalized Stokes model (binned according to the data), and $q$ and $u$ correspond to the normalized IXPE Stokes data, with $\sigma_q$ and $\sigma_u$ the associated data errors. Equation \ref{eq:model_nstokes} depends on Equations \ref{eqn:Ix}, \ref{eqn:Qx} and \ref{eqn:Ux}, for which the pulsar model $I_\CX$ is folded according to IXPE response functions (RMF and ARF), including approximately the effects of ISM absorption \citep{Wilms2000} with a fixed value $N_\mathrm{H}=3.27\times 10^{21}\mathrm{cm}^{-2}$ \citep{weisskopf11}.  
The  $I_\mathrm{\Xband}$ Stokes is fitted separately and the results for the coefficients $\alpha$ and $b_\Istk$ are summarized in Table \ref{tab:table1}. 

In order to build posterior distributions of the best fitted parameters, we use an MCMC analysis \citep{Foreman-Mackey2013}, which also helps us to verify potential multiple local minima and choose the best space of parameter solutions. By visual inspection of the data, it is already clear that a large phase-shift is not present in the polarization angle at the main pulse. Therefore, we set hard boundaries to search for a phase shift $\delta_\mathrm{\Pulse}$ in the range $[-5,5\%]$. As for the interpulse, visual inspection does not provide clear evidence for or against a phase-shift, prompting us to allow broader boundaries for $\delta_\mathrm{\IPulse}$   in the range $[-30,30]\%$.  In the following, for all MCMC analyses we build posterior distributions using 100 walkers and 10,000 steps, discarding the initial $20\%$ of iterative steps.  

\section{Discussion and conclusions}
\label{sec:discussion_conclusions}

Figure \ref{fig:binned_Obs1}, \ref{fig:binned_Obs2} and \ref{fig:binned_Obs3} show the best-fitted model to the data for each of the three sets of IXPE observations. In all cases, the model provides a satisfactory description of the data for almost all the rotational phases of the pulsar, except for the phases around the main pulse $\approx 0.95-1.00$. In this specific phase range, there is a deviation between the model and the data, particularly evident in the normalized Stokes $Q/I$ for the third observations.  This deviation propagates to the polarization angle, which shows a relatively large swing that our model is unable to reproduce accurately. Nevertheless, it is important to note that this minor discrepancy does not impact the overall results and demonstrates, for the first time, that a simple linear transformation of the Stokes parameters from the optical to the X-ray band is sufficient to explain the IXPE observations of the Crab pulsar. This suggests that a common mechanism is, therefore, likely responsible for the emission in the optical and X-ray bands.  This result provides further observational confirmation for previous theoretical studies based on synchrotron radiation, which provide a fairly good description of the optical pulsed emission and polarization \citep[see e.g.,][]{Petri2005}, as well as the SED between optical and X-ray of Crab pulsar \citep[see e.g.,][]{Harding2021}.

Our results represent a step forward compared to the recent analysis of Crab by \citet{Wong2024}. They applied a simultaneous fitting technique \citep{Wong2023} to the IXPE data, a sensitive method based on Chandra X-ray observations to model the pulsar (and nebula), enabling them to characterize the phase-dependent polarization properties in X-rays. Then, they compared the phase-dependent polarization properties in the optical \citep{slowikowska09} with those found in X-ray, reporting a substantial difference between the two. This led them to conclude, contrary to our results, that different emission mechanisms (or sites) are responsible for the optical and X-ray emissions of the Crab pulsar.
However, unlike our work, they did not report further attempts to establish a relationship between the optical and X-ray phase-dependent polarization properties.

From our phenomenological model (Equations \ref{eqn:Ix}, \ref{eqn:Qx} and \ref{eqn:Ux}) and the results presented in Table~\ref{tab:table1}, we can see that if we neglect the nebula contribution, i.e., setting $b_\Istk = b_\Qstk = b_\Ustk =0$, then the (phase-dependent) polarization degree for the pulsar alone can be simply written as $p =  p_\Vband\, \beta/\alpha \approx 0.46 p_\Vband$. This implies that in the IXPE band, Crab pulsar is about $54\%$ less polarized than in the optical band, suggesting an underlying energy-dependent polarized emission. Similarly, without nebula contribution, the polarization angle for the pulsar in the IXPE band is found to be similar to that in the optical band but with the inclusion of a phase shift at the interpulse that is discussed below. On the other hand, we also obtain that the polarization degree for the nebula component without the pulsar contribution is  $\sqrt{b_\Qstk^2 + b_\Ustk^2}/b_\Istk \approx 22\% $, while the polarization angle is $(1/2)\arctan(b_\Ustk/b_\Qstk)\approx 135^\circ$ as also clearly evident in Figure \ref{fig:binned_Obs1}, \ref{fig:binned_Obs2}, and \ref{fig:binned_Obs3}.

Notably, by applying the model mentioned above,  we also found a variable phase-shift at the interpulse $\delta_\mathrm{\IPulse}=7.42^{+5.94}_{-4.42}$, $21.23^{+3.52}_{-2.74}$, and $-2.16^{+9.13}_{-13.88}\%$  ($\approx2\sigma$, $8\sigma$ and within $1\sigma$ away from zero, respectively) when performing separately\footnote{By combining all three observations, it is not possible to obtain a unique measurement of the phase shift at the interpulse as the associated posterior distribution shows multiple peaks.} the MCMC analysis for the three sets of IXPE observations. (Notice that $\delta_\mathrm{\IPulse}$ for the first observation is nearly $2\sigma$ away from zero, but also with a bi-modal distribution, as shown in the posterior distribution, with a secondary solution around $\delta_\mathrm{\IPulse}\sim12\%$.) Instead,  phase-shifts marginally consistent with zero (less than $2
\sigma$ away from zero) are still present at the main pulse for the different observations:  $\delta_\mathrm{\Pulse}=-0.67^{+0.51}_{-0.63}$, $-0.90^{+0.57}_{-0.70}$, $-1.13^{+1.61}_{-2.24}\%$  (see also Table~\ref{tab:table1} and  posterior distributions in Figure~\ref{fig:shift_corner_obs1}, \ref{fig:shift_corner_obs2}, and \ref{fig:shift_corner_obs3}). 
Furthermore, including the phase-shift in our model improves the reduced $\chi^2$ statistic compared to the results obtained without accounting for phase-shifts, as also shown in Table~\ref{tab:table1}. This statistical improvement\footnote{From the probability density function for the $\chi^2$ distribution, we obtain that  the odds ratio between the models including and excluding the phase shifts is  $\approx2$.} indicates that the inclusion of phase-shifts in our analysis is indeed necessary to properly explain the IXPE data. The phase shifts $\delta_\mathrm{\IPulse}$ for the first two IXPE observations are substantially larger than the difference in the location of the interpulse peak between optical and X-rays, which is smaller than $1\%$, ruling it out as the main contribution to the measured $\delta_\mathrm{\IPulse}$.  On the other hand, we are aware that dead time can produce a deformation of the light curve and possibly a shift in the peak phase, but on the basis of simulations it is much less than $1\%$ (namely 330\,$\mu$s), and certainly much lower than $\delta_\mathrm{\IPulse}$ found in our analysis for the first two IXPE observations. A further discussion on dead time and its negligible effects on IXPE observations of Crab pulsar can also be found in \cite{Bucciantini2023}.

We also explored the scenario in which the optical Stokes parameters used in our phenomenological model might be affected by contamination from the PWN Knot. By removing $90\%$ from the DC component from the optical Stokes parameters, we redo the analysis, reporting the best fitted parameters at the end of Table~\ref{tab:table1} and showing the corresponding phase-dependent polarization curves in X-ray in Figure~\ref{fig:binned_DC}. If we neglect the X-ray nebula contribution, we now obtain $p =  p_\Vband\,\beta/\alpha \approx 0.56 p_\Vband$ for pure X-ray pulsar emission. 
%Given the unknown optical PWN Knot contribution to our model, we conclude that the polarization degree for the pulsar is reduced by a factor $\approx (0.46 - 0.56)$ compared to the optical band. 
By neglecting or accounting for optical Knot contribution to our model, we conclude that the polarization degree for the pulsar is reduced by a factor $\approx (0.46 - 0.56)$ compared to the optical band \citep[for further discussion on the Knot emission in the optical, see][]{moran13}. 
Other results remain fairly consistent with the analysis discussed above for all three IXPE observations, with exception of the interpulse for the second IXPE observation where even a larger phase shift is present $\delta_\mathrm{\IPulse}=32.91^{+1.23}_{-1.51}\%$. Consequently, determining the final estimate remains somewhat uncertain; so we consider this measurement as a lower limit $\delta_\mathrm{\IPulse} > 18.49\%$ (for the second observation). However, it is worth noting that all error intervals in the fitted parameters reported in Table~\ref{tab:table1} can be further reduced using an unbinned likelihood analysis \citep[][see also \citealt{marshall21}]{GonzalezCaniulef2023,heyl24}, but this is left for future work.

By neglecting the phase shifts and the nebula contribution in our phenomenological model, the entire phase-dependent polarization angle for the pulsar results the same in both the optical and X-ray bands, i.e., $\psi(\phi,\delta_\mathrm{\IPulse}=0, \delta_\mathrm{\Pulse}=0) = \varphi_\Xband(\phi,\delta_\mathrm{\IPulse}=0, \delta_\mathrm{\Pulse}=0)=\psi_\Vband(\phi)$. Figure \ref{fig:loop} shows colour-coded Stokes parameter $Q$ and $U$ as a vector diagram, considering the pulsar alone extracted from our phenomenological model.
The loops in the plane $Q$-$U$  look similar to those shown in Figure 4 in \cite{slowikowska09}, but with smaller amplitude due to the reduced polarization degree in X-rays, as discussed above,  by $\approx (0.46 - 0.56)$, compared to the optical band.  While a first-principles theoretical model that matches these $Q$-$U$ loops remains to be developed, significant progress has been obtained with particle-in-cell simulations of pulsar magnetospheres, which partially explain the $Q$-$U$ loops observed in the optical band \citep[see e.g.,][]{Cerutti2016}

In presence of vacuum birefrigence, a highly magnetized and fast rotation pulsar may produce phase-shift in the polarization angle between different bands, e.g., with the polarization angle swing in the optical band lagging behind the X-ray band.  The large phase-shifts at the interpulse for the first two IXPE observations are unlikely to be associated to vacuum birefringence, as they are about one order-of-magnitude larger than early theoretical calculations that predicted values in the range $\delta \approx 1-2\%$ for a corrotating dipole model or Deutch model \citep{heyl00}.  Instead, our results suggest some form of variability in the polarization angle swing at the interpulse, as observed when analyzing each IXPE observation separately. The origin of this variability is unknown. A theoretical study of this phenomenon is outside the scope of this paper and will be left for future work. 

Nevertheless, besides producing phase-shifts in the polarization angle, vacuum birefringence primarily causes the polarization modes of the photons readapt to the local magnetic field as radiation propagates through the magnetosphere. This effect takes place up to the polarization limiting radius, located at several tens of NS radii; beyond that the magnetic field weakens and the photon polarization mode freezes \citep{heyl02}. The non-detection of a (positive) phase shift at the main pulse in all three IXPE observations likely indicates that the emission mechanism takes place beyond the polarization limiting radius, perhaps outside the light cylinder as discussed in several current sheet models/simulations of pulsar magnetosphere \citep[see e.g.,][]{Petri2005, Contopoulos2010, Cerutti2016,Harding2017}. On the other hand, the variability of the phase-shift at the interpulse suggests that the emission mechanism might be located in regions where the pulsar magnetosphere might undergo through some form of rearrangement on time scales comparable to that between different IXPE observations. However, this scenario is in conflict with the X-ray pulse profile of Crab pulsar, which has been studied for many years and shows no variability at the interpulse. If the variability of the polarization angle swing at the interpulse is confirmed by future polarization observations, it would suggest some form of decoupling between the flux and formation of the polarization angle. The origin of this phenomenon will pose a new challenge for theoretical studies of pulsar emission and polarization.

In addition, another consequence of vacuum birefringence is that, at the polarization limiting radius,  the polarization modes of the radiation align with the more uniform projected magnetic field on the plane in the sky, as well as with the projected magnetic axis \citep{heyl02}.  As the pulsar rotates, therefore, the phase-dependent polarization angle should follow a simple rotating vector model \citep{Radhakrishnan1969,GonzalezCaniulef2023}. However, Crab pulsar behaves differently, deviating from the rotation vector model in both optical and X-rays polarimetric observations.   This deviation hints again that the emission mechanisms for the main pulse and interpulse are likely produced beyond the polarization limiting radius, and remain unaffected by vacuum birefringence.

If the X-ray emission indeed takes place far away in the magnetosphere, its reduced polarization compared to the optical band might indicate that these emissions originate from different regions. Specifically, while the X-rays might be produced close to the light cylinder radius, where turbulence due to reconnection at the Y-point might be stronger and hence the level of polarization is relatively low, the optical emission might be located further out, in the striped wind, where turbulence might have decayed, allowing for a higher degree of polarization. For emission located at of beyond the Light Cylinder the expected phase lag due to different location of the emitting region is typically very small \citep{Petri13a}.  This demonstrates that X-ray polarization information may serve to probe pulsar magnetospheres and potentially the current sheet scenario discussed in particle-in-cell simulations \citep[for a recent review see e.g.,][]{Philippov2022}.  

Our results pave the way for a systematic search, via optical and X-ray polarimetric observations, for phase-shifts in sources similar to the Crab pulsar, as well as for gaining insight into the emission mechanism. In particular, PSR B0540+69 is a promising source. Located in the Large Magellanic Cloud, 
it is considered the ``Crab Twin''. This source shows phase-aligned light curves from radio to gamma rays, as well as dominant non-thermal emission at optical and X-rays. Polarimetric observations with HST/WFP detected phase-averaged $\mathrm{PD}=16\pm4\%$ and $\mathrm{PA} = 22^\circ\pm12^\circ$ \citep{mignani2010}. Recent IXPE observations revealed, for a phase-dependent analysis,  two bins with $\mathrm{PD}$ of $68\pm20\%$ and $62\pm20\%$ \citep{xie2024}. Further polarimetric observations in both the optical and X-ray bands are needed to verify whether common phase-dependent polarization properties are also present in this source.

Expanding the comparison to include gamma-ray observations of the Crab should be also a key objective for future research \citep[see e.g.,][]{Dean2008, Li2022}.

\begin{acknowledgements}
We thank the referee for their constructive comments. We also thank Benjamin Crinquand for discussions that benefited this paper.
DGC acknowledges support from a CNES fellowship. JH acknowledges support from the Natural Sciences and Engineering Research Council of Canada (NSERC) through a Discovery Grant and the Canadian Space Agency through the co-investigator grant program. NB was supported by the INAF MiniGrant ``PWNnumpol - Numerical Studies of Pulsar Wind Nebulae in The Light of IXPE''.
F.X. is supported by National Natural Science Foundation of China (grant No. 12373041), and special funding for Guangxi distinguished professors (Bagui Xuezhe).
This work is partially supported by MAECI with grant CN24GR08 
``GRBAXP: Guangxi-Rome Bilateral Agreement for X-ray Polarimetry in 
Astrophysics''.
This research was enabled in part by support provided by Compute Canada (www.computecanada.ca), UBC ARC Sockeye infrastructure, and the SciServer science platform (www.sciserver.org). 
\end{acknowledgements}


\bibliographystyle{aa}
\bibliography{main}

\appendix 
\section{Optical Stokes}
\label{appendix:optical}
\begin{figure}
    \centering
    \includegraphics[width=\linewidth]{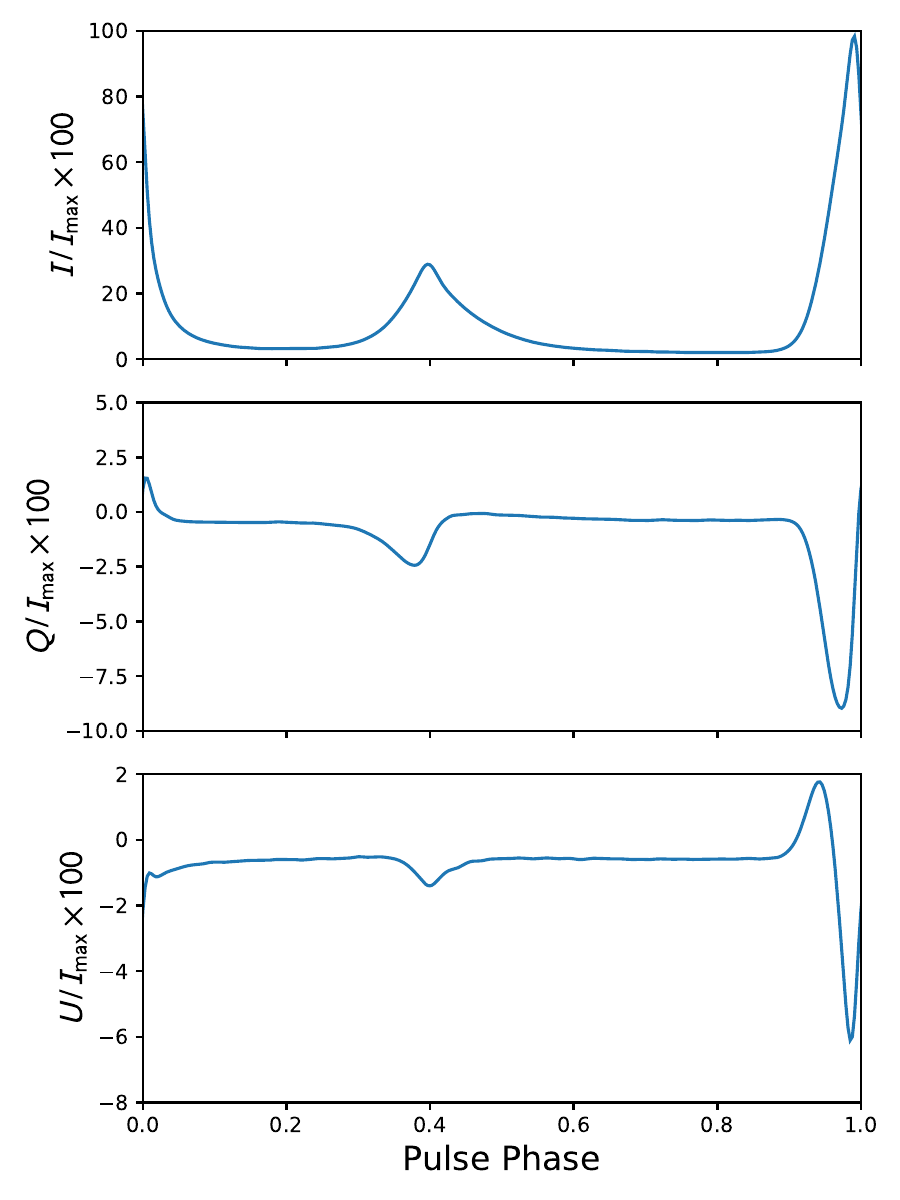}
    \caption{Optical Stokes parameters for Crab pulsar ($I_\Vband, Q_\Vband, U_\Vband$).}
    \label{fig:optical}
\end{figure}

Figure \ref{fig:optical} shows the Stokes parameters in the optical band ($I_\Vband, Q_\Vband, U_\Vband$), which are taken from \citet{slowikowska09} and smoothed with a Radial basis function interpolation. The interpulse and main peaks are located at 0.396 and 0.993 in phase, respectively, relative to the radio peak.

\section{Intensity from Chandra X-ray observations}
\label{appendix:xray}
\begin{figure}
    \centering
    \includegraphics[width=\linewidth]{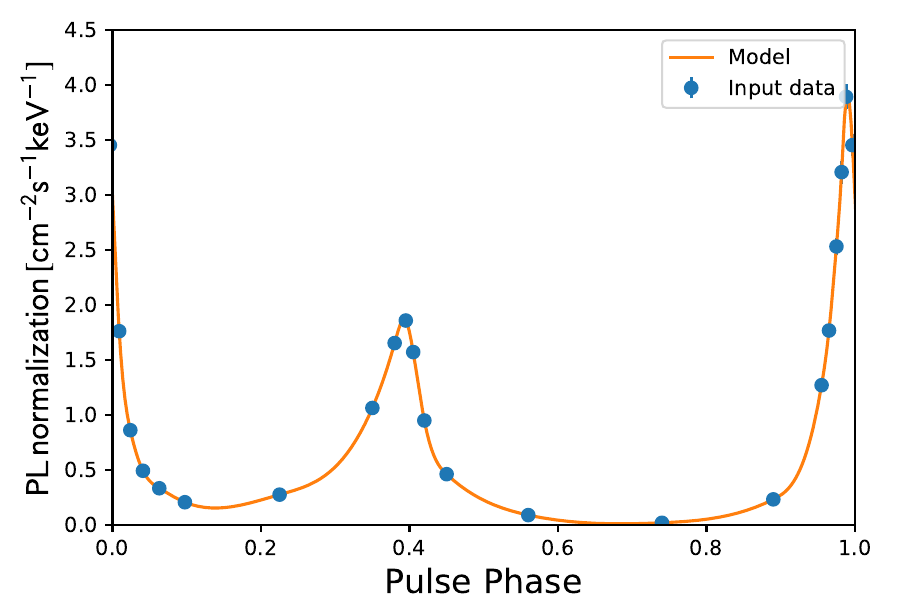}
    \includegraphics[width=\linewidth]{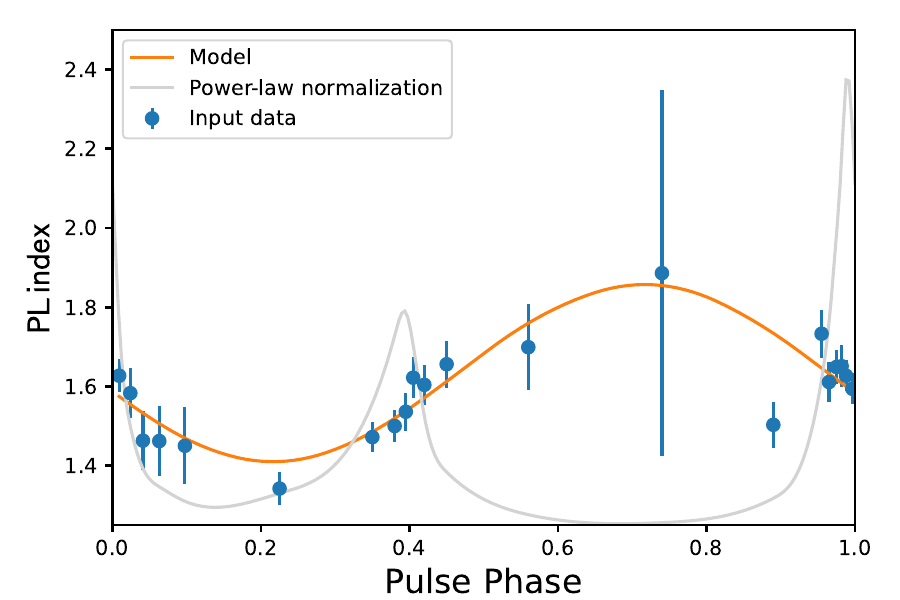}
    \caption{Power-law normalization and index for the phase-dependent spectrum of Crab pulsar in soft X-rays. Data points (in blue) are taken from \citet{weisskopf11}. The orange curve in the first and second panel  correspond to a spline interpolation and a sinusoidal fit, respectively.}
    \label{fig:chandra}
\end{figure}

Figure \ref{fig:chandra} shows the soft X-ray spectrum of Crab pulsar used to model $I_\CX$, based on Chandra observations from \citet{weisskopf11}, which is parameterized as a power law whose
normalization and index vary in phase. The spectral index as a function of the pulse phase has been fitted to a sinusoidal in order to have a sufficiently smooth pulse profile. The interpulse and main peaks are located at 0.393 and 0.990 in phase, respectively, relative to the radio peak.

\section{Fits for the three observations of Crab Pulsar by IXPE}
\label{appendix:all_obs}

\begin{figure*}
    \centering
    \includegraphics[width=0.75\linewidth]{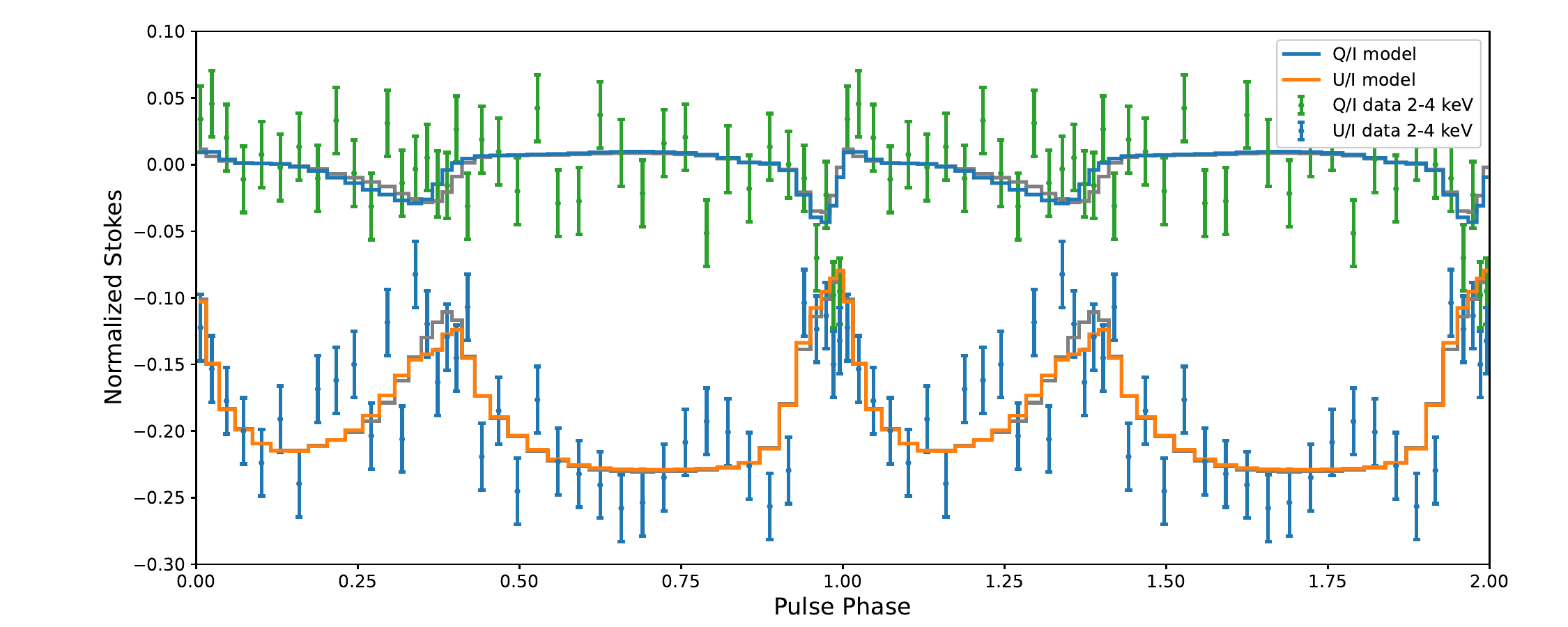}
    \includegraphics[width=0.75\linewidth]{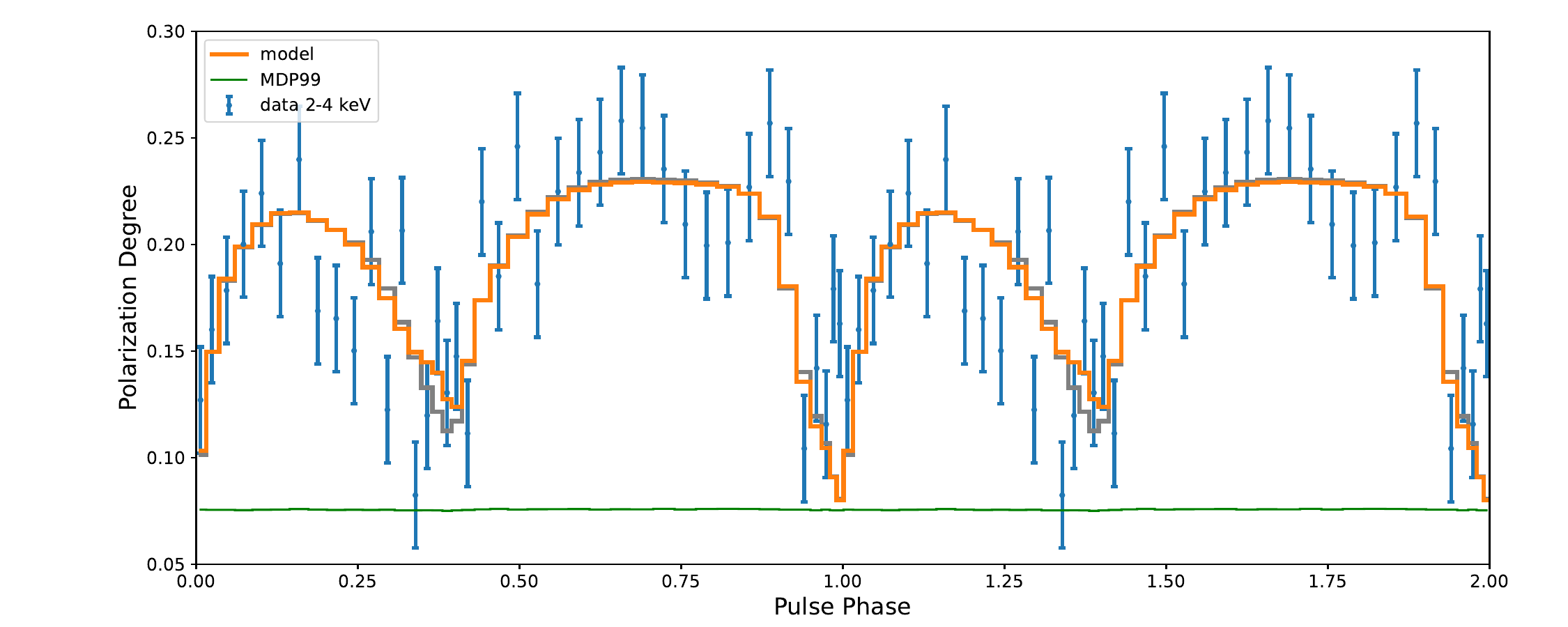}
    \includegraphics[width=0.75\linewidth]{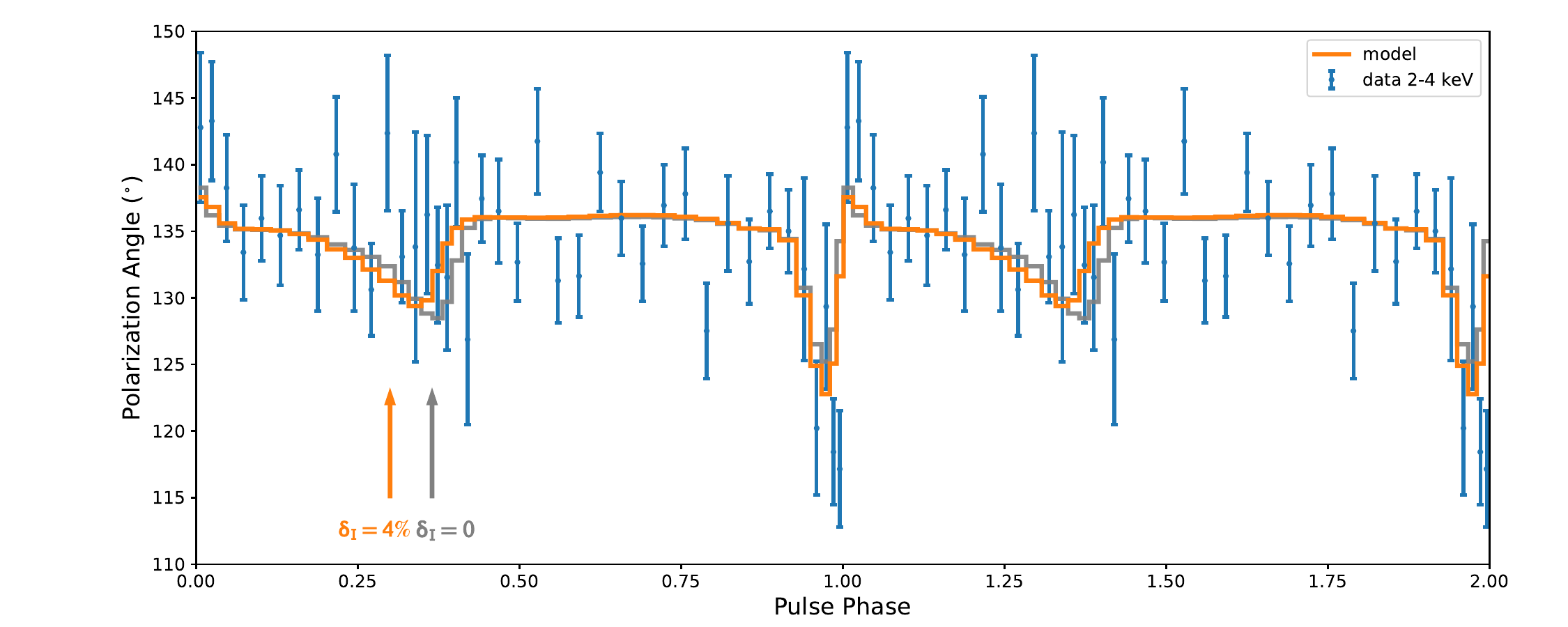}
    \includegraphics[width=0.75\linewidth]{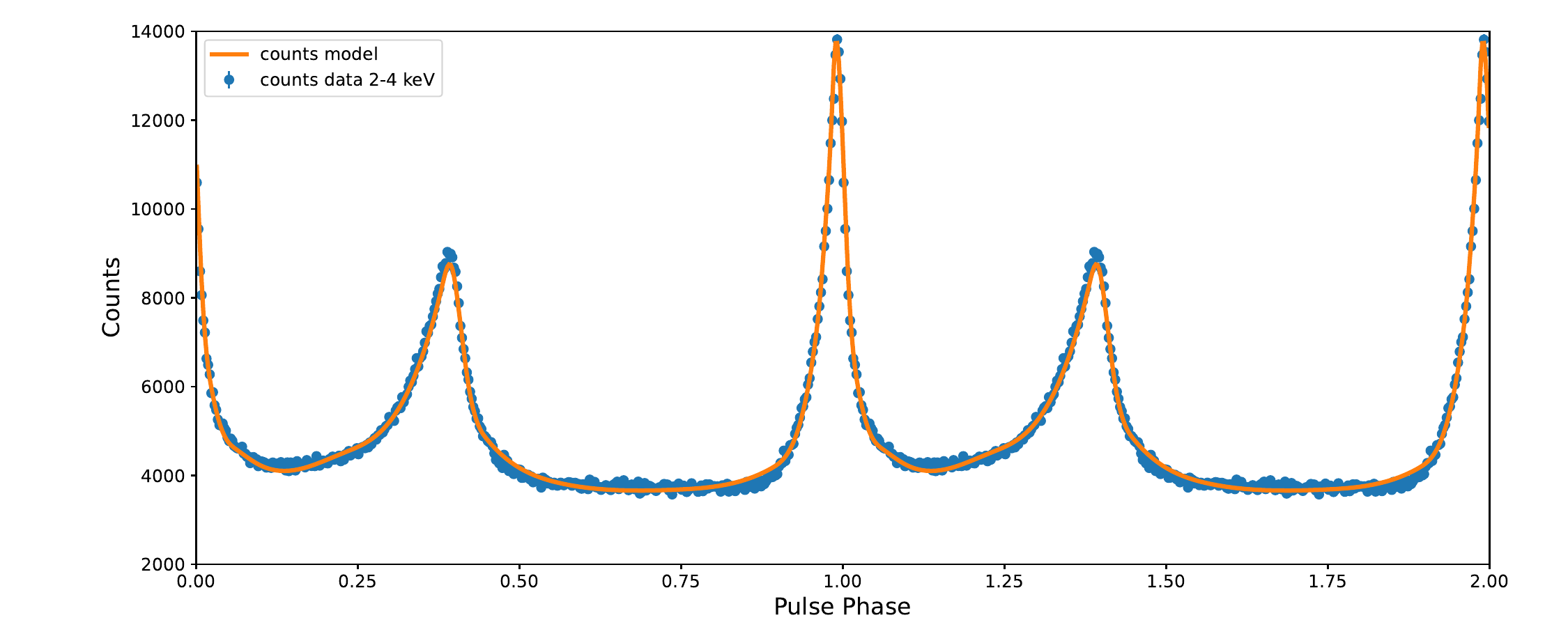}
    \caption{Best fitted model to the first observation of Crab pulsar by IXPE in the $2-4$~keV range. Data reduction is performed with \textsc{ixpeobssim} package considering equi-populated binning for the Stokes parameters (first panel), as well as for the polarization degree (second panel) and polarization angle (third panel).  The green line in the second panel corresponds to $\mathrm{MDP}_{99}$. The gray solid lines in the first three panels correspond to the best-fitted model without phase shift. The orange and gray arrows highlight the polarization angle swing, at the interpulse,  with and without a phase shift, respectively.  For completeness, we include the pulse profile with 400 equi-spaced bins (fourth panel). 
%     For the latter, the orange curve shows the fitted phenomenological model based on Chandra pulse profile \citep{weisskopf11}, which includes nebula contribution to match the IXPE observation. 
    }
    \label{fig:binned_Obs1}
\end{figure*}

\begin{figure*}
    \centering
    \includegraphics[width=\linewidth]{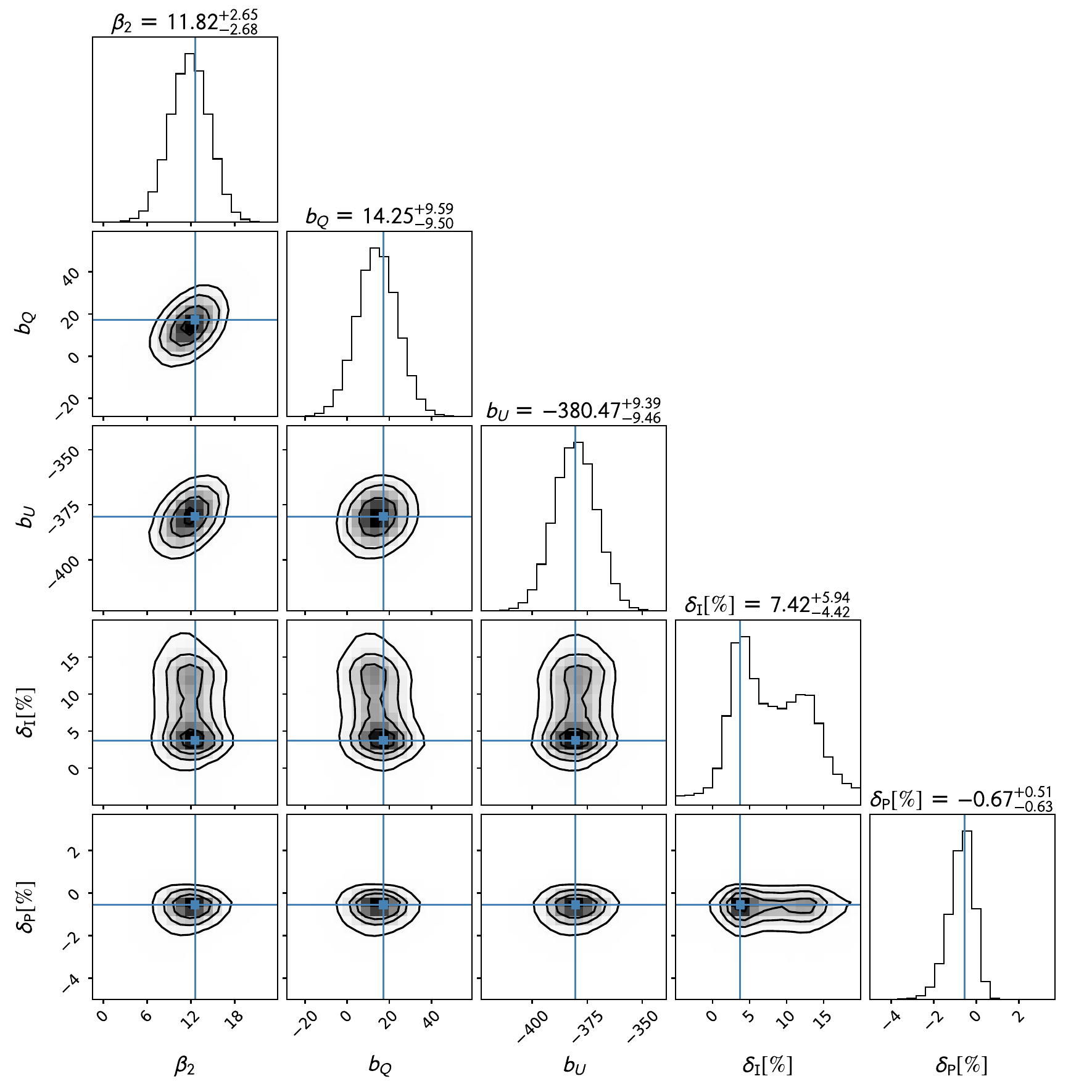}
    \caption{Posterior distributions for the first observation of Crab pulsar by IXPE. The model parameters include the phase-shifts at the main pulse ($\delta_\mathrm{P}$) and interpulse ($\delta_\mathrm{I}$). The MCMC analysis is performed using 100 walkers and 10,000 steps. The blue lines show the solution obtained with a minimization routine. 
    }
    \label{fig:shift_corner_obs1}
\end{figure*}

\begin{figure*}
    \centering
    \includegraphics[width=0.8\linewidth]{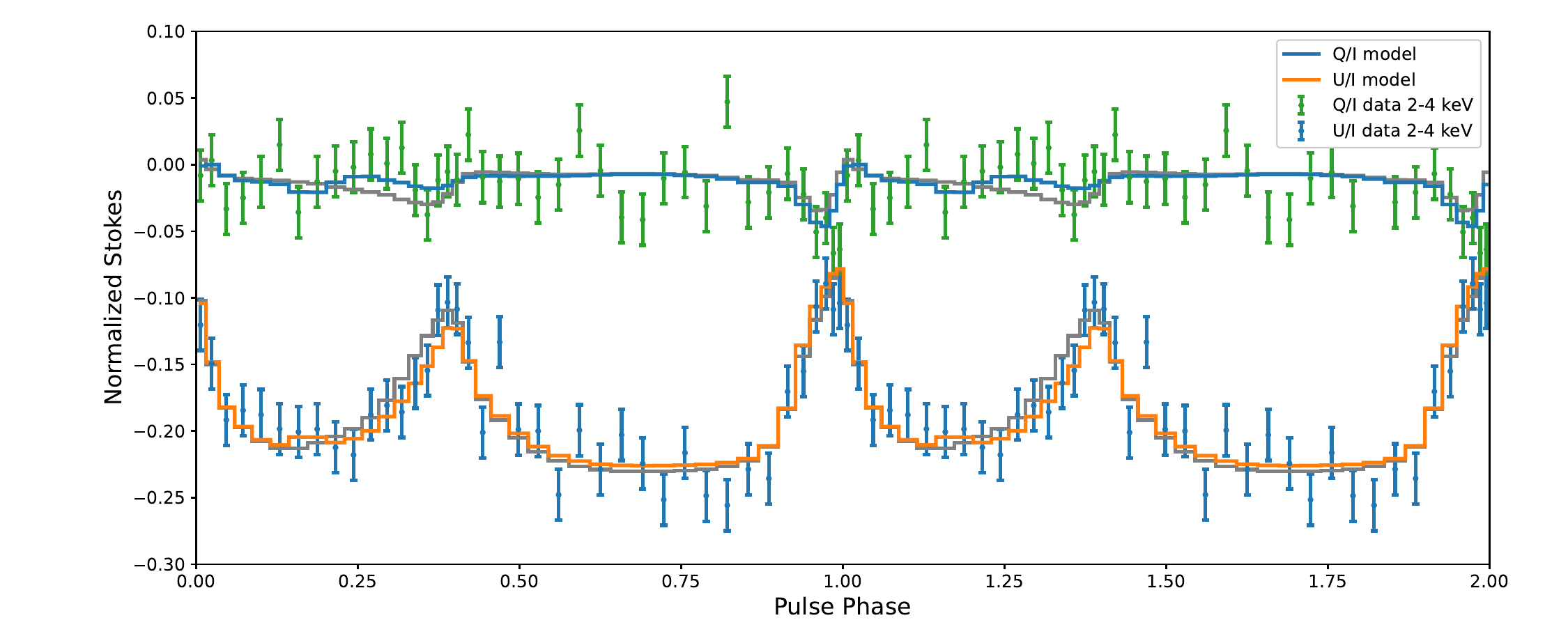}
    \includegraphics[width=0.8\linewidth]{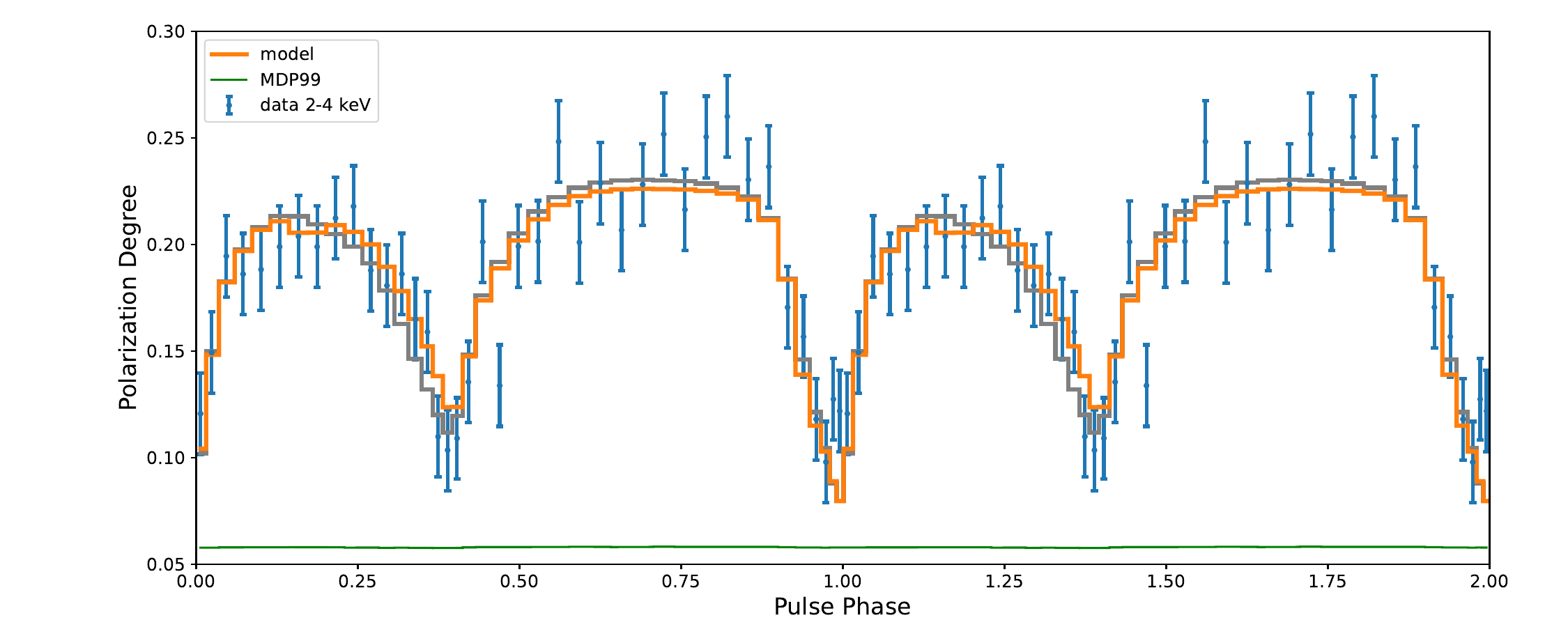}
    \includegraphics[width=0.8\linewidth]{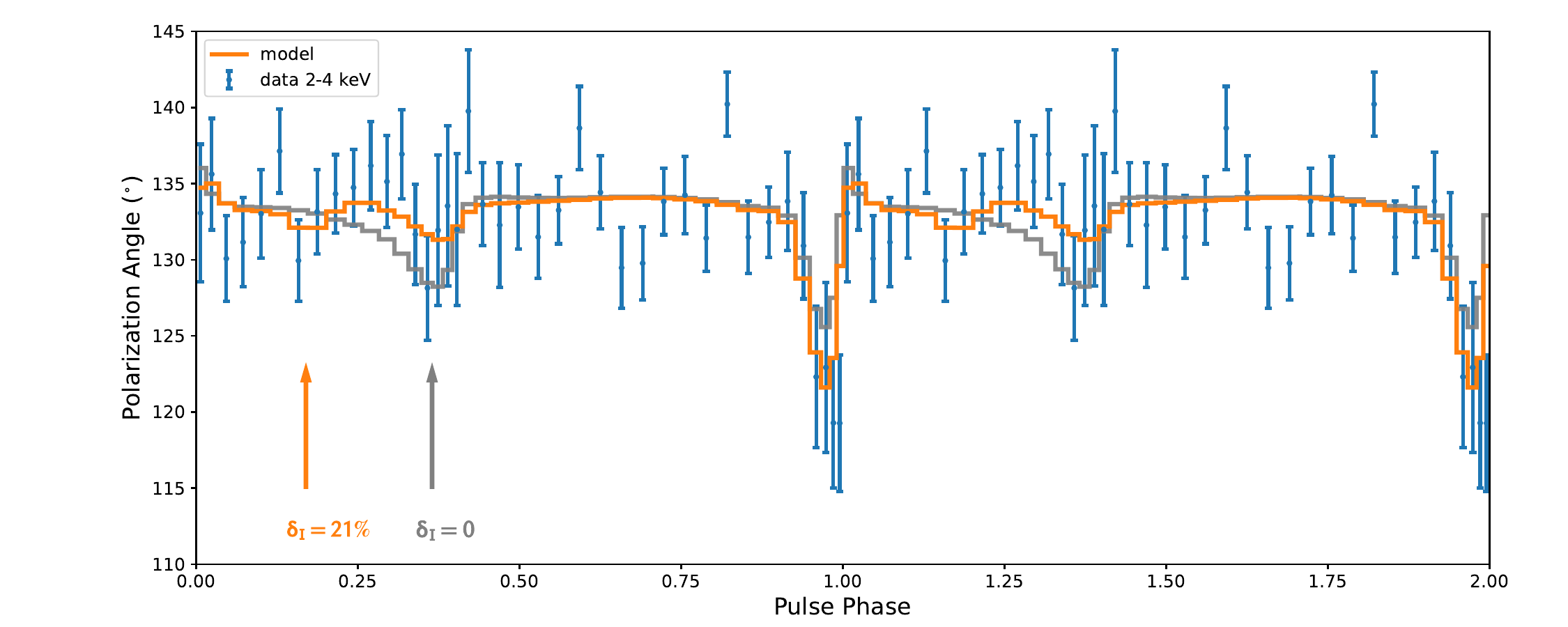}
    \includegraphics[width=0.8\linewidth]{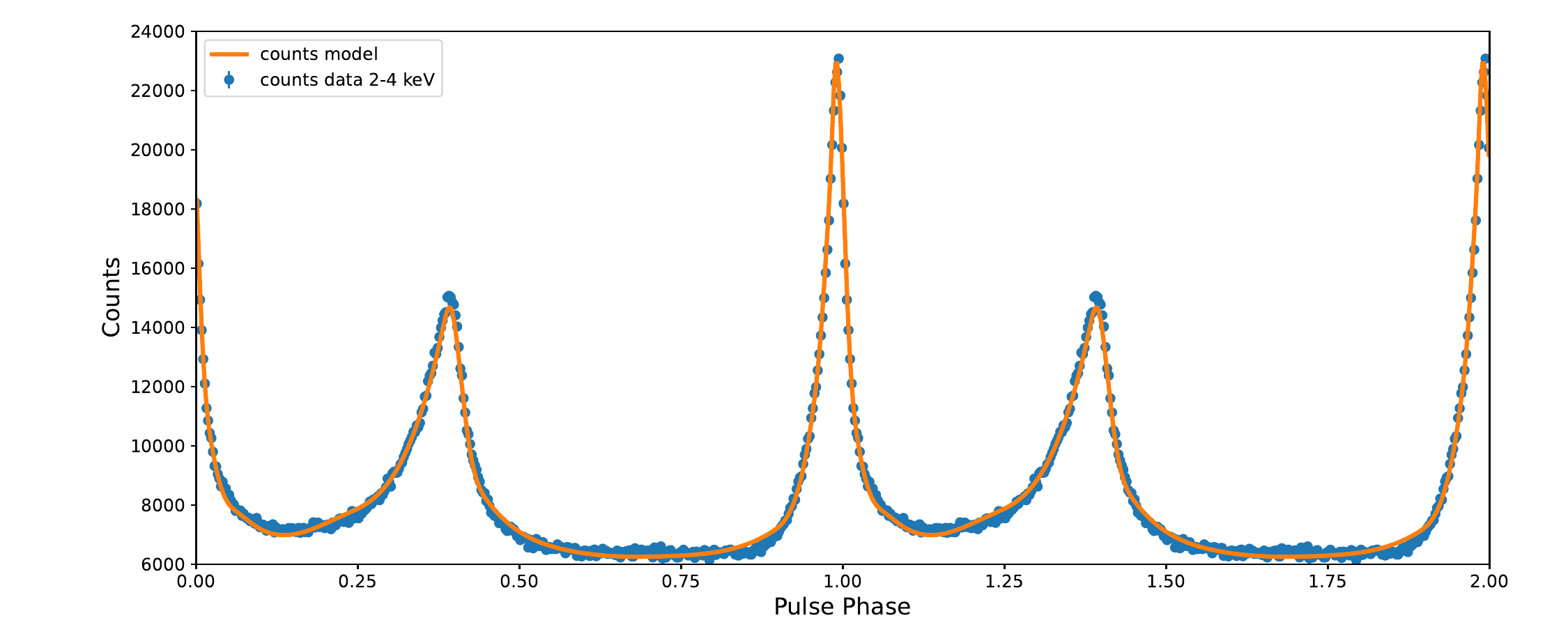}
    \caption{Same as Figure \ref{fig:binned_Obs1} for the second observation of Crab pulsar by IXPE.}
    \label{fig:binned_Obs2}
\end{figure*}

\begin{figure*}
    \centering
    \includegraphics[width=\linewidth]{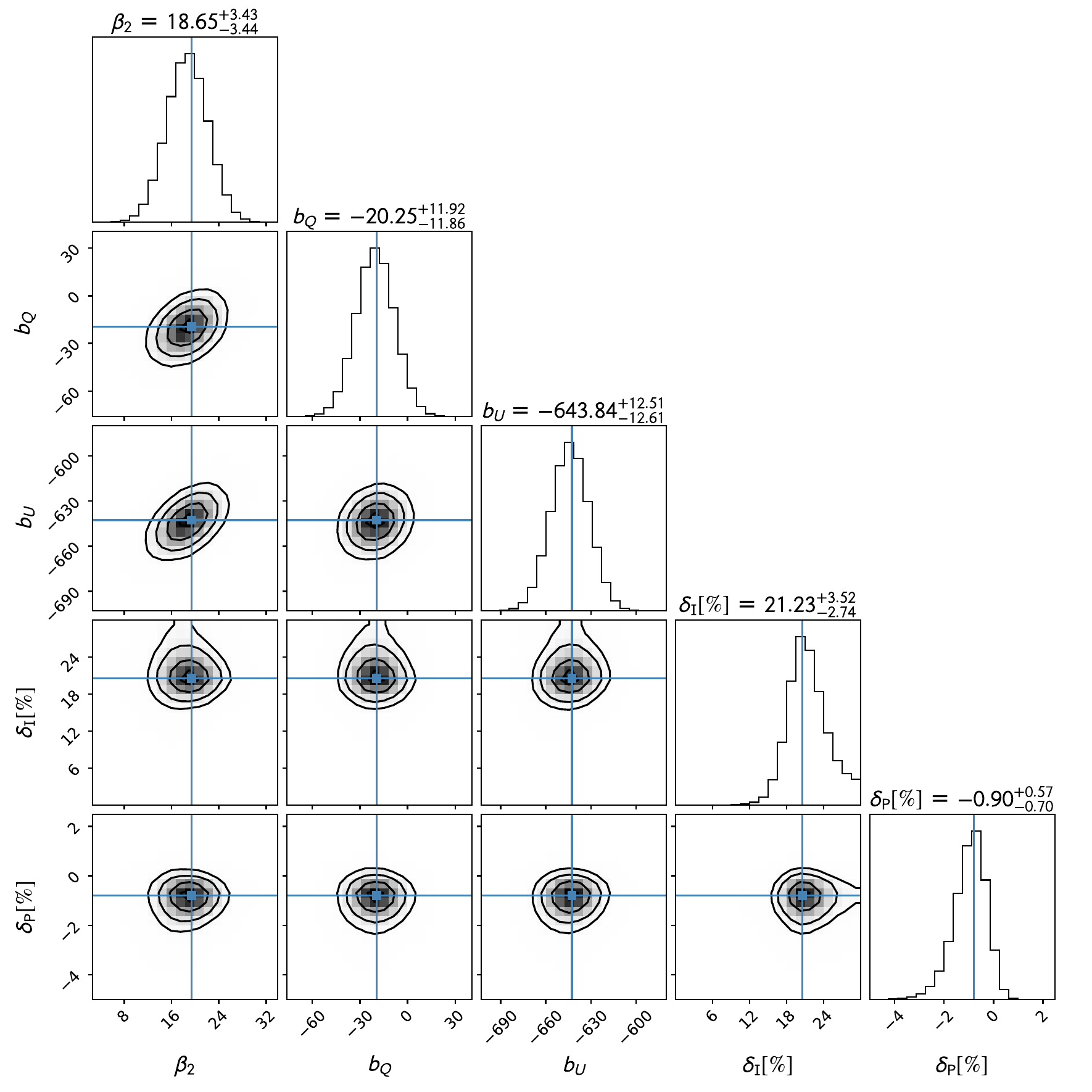}
    \caption{Same as Figure \ref{fig:shift_corner_obs1} for the second IXPE observation of Crab pulsar.
    }
    \label{fig:shift_corner_obs2}
\end{figure*}

\begin{figure*}
    \centering
    \includegraphics[width=0.8\linewidth]{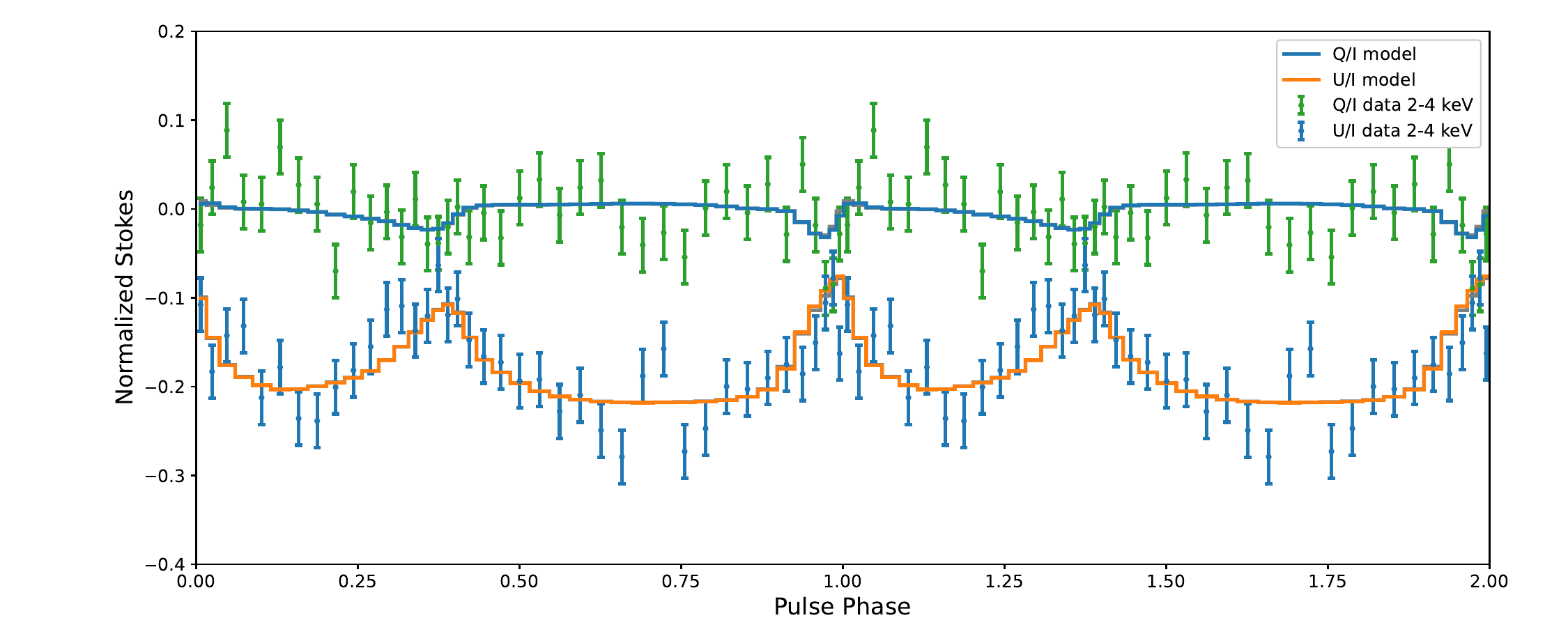}
    \includegraphics[width=0.8\linewidth]{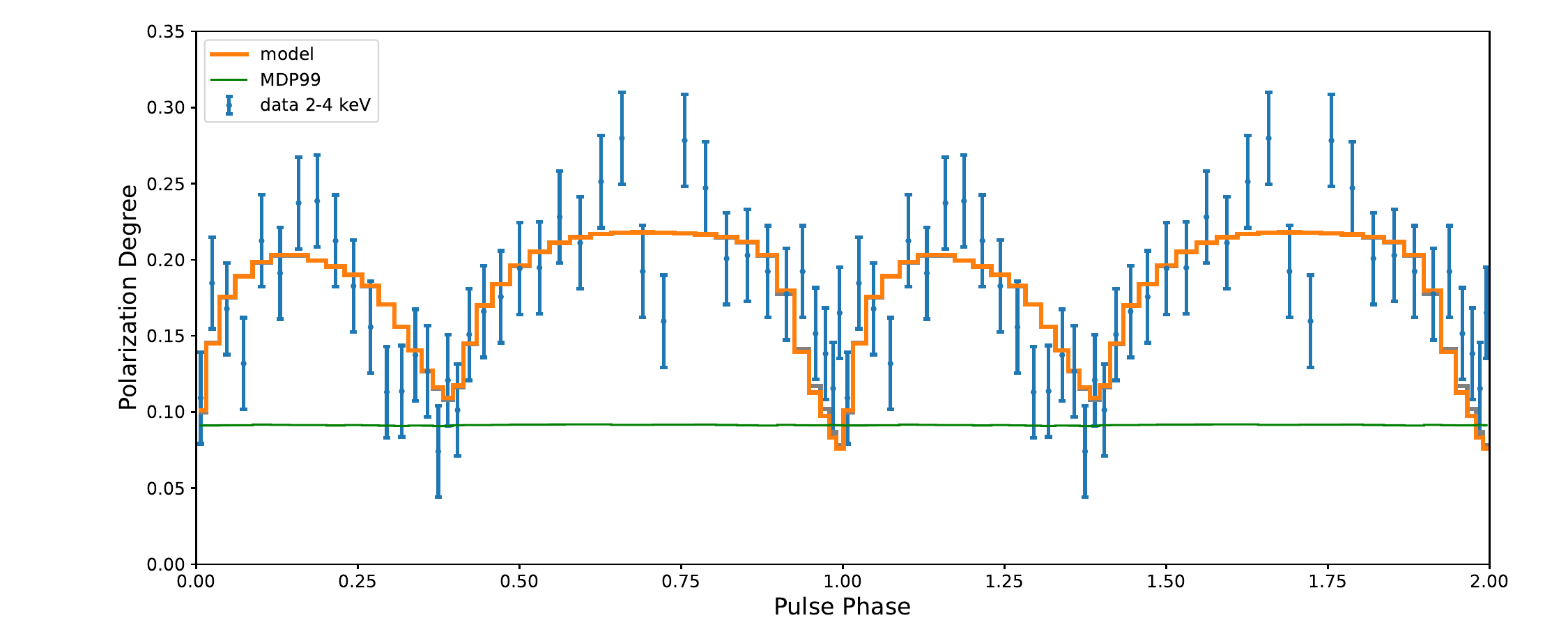}
    \includegraphics[width=0.8\linewidth]{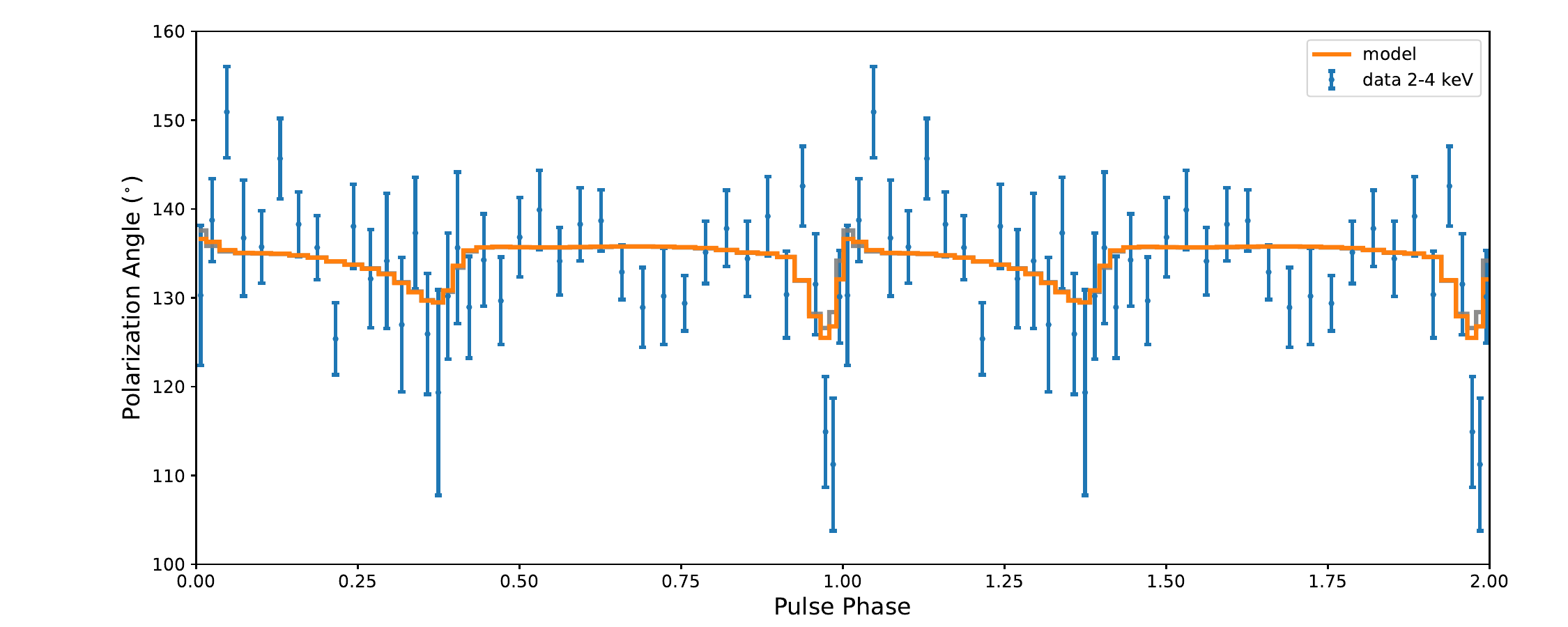}
    \includegraphics[width=0.8\linewidth]{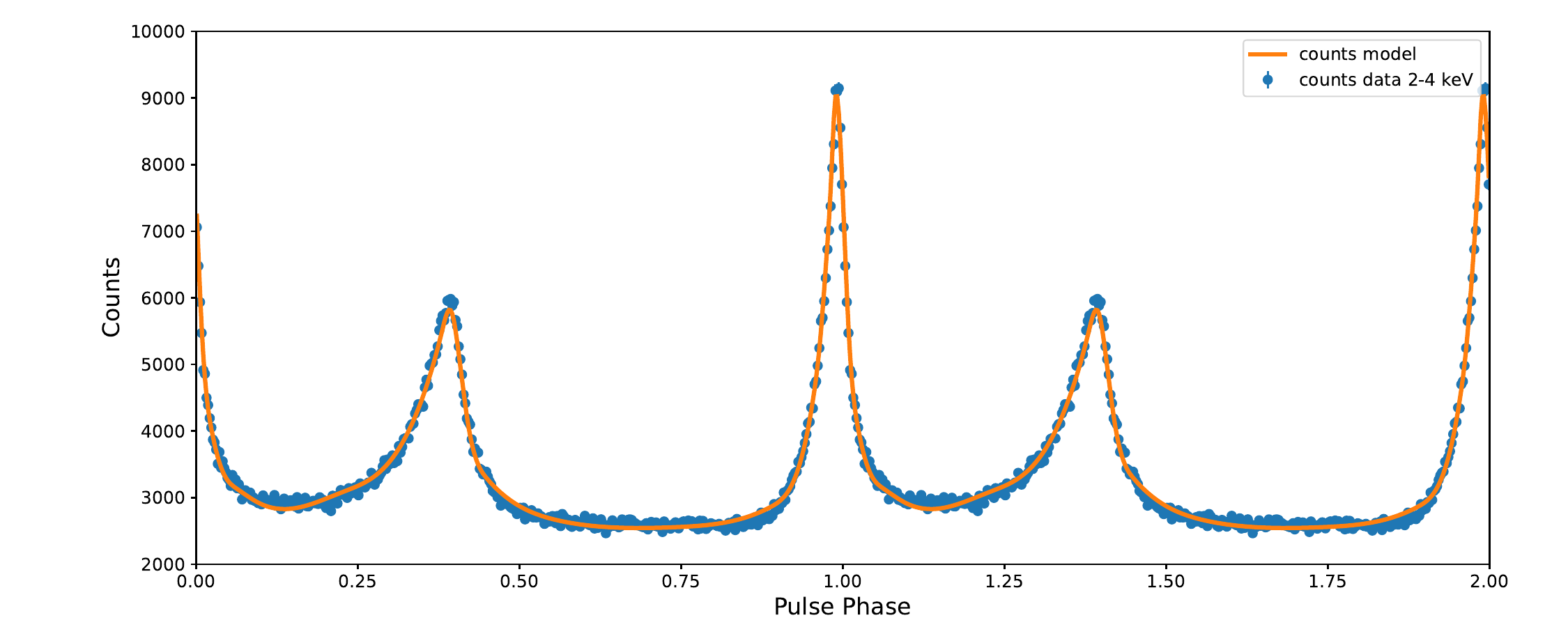}
    \caption{Same as Figure \ref{fig:binned_Obs1} for the third observation of Crab pulsar by IXPE. 
    }
    \label{fig:binned_Obs3}
\end{figure*}

\begin{figure*}
    \centering
    \includegraphics[width=\linewidth]{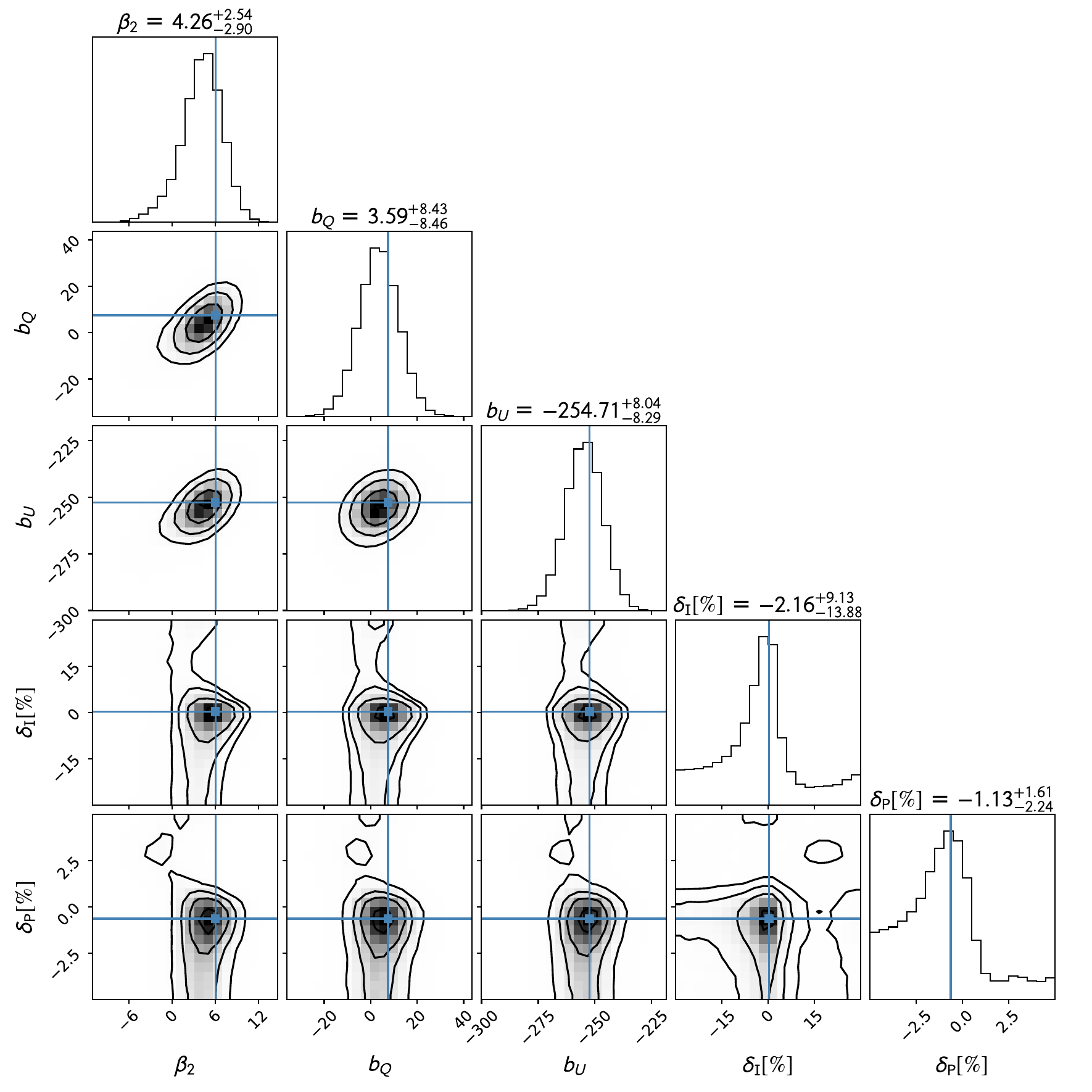}
    \caption{Same as Figure \ref{fig:shift_corner_obs1} for the third observation of Crab pulsar by IXPE.
    }
    \label{fig:shift_corner_obs3}
\end{figure*}

\begin{figure*}
    \centering
    \includegraphics[width=0.52\linewidth]{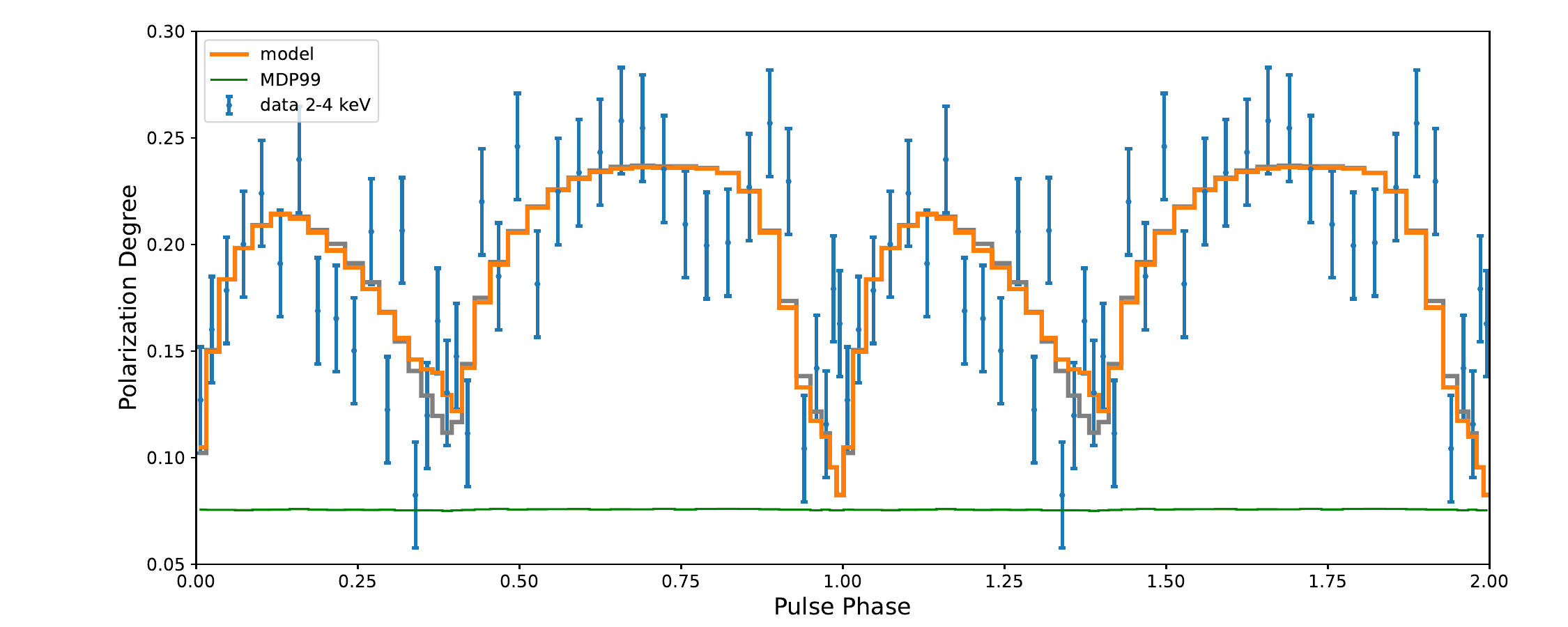}
    \includegraphics[width=0.52\linewidth]{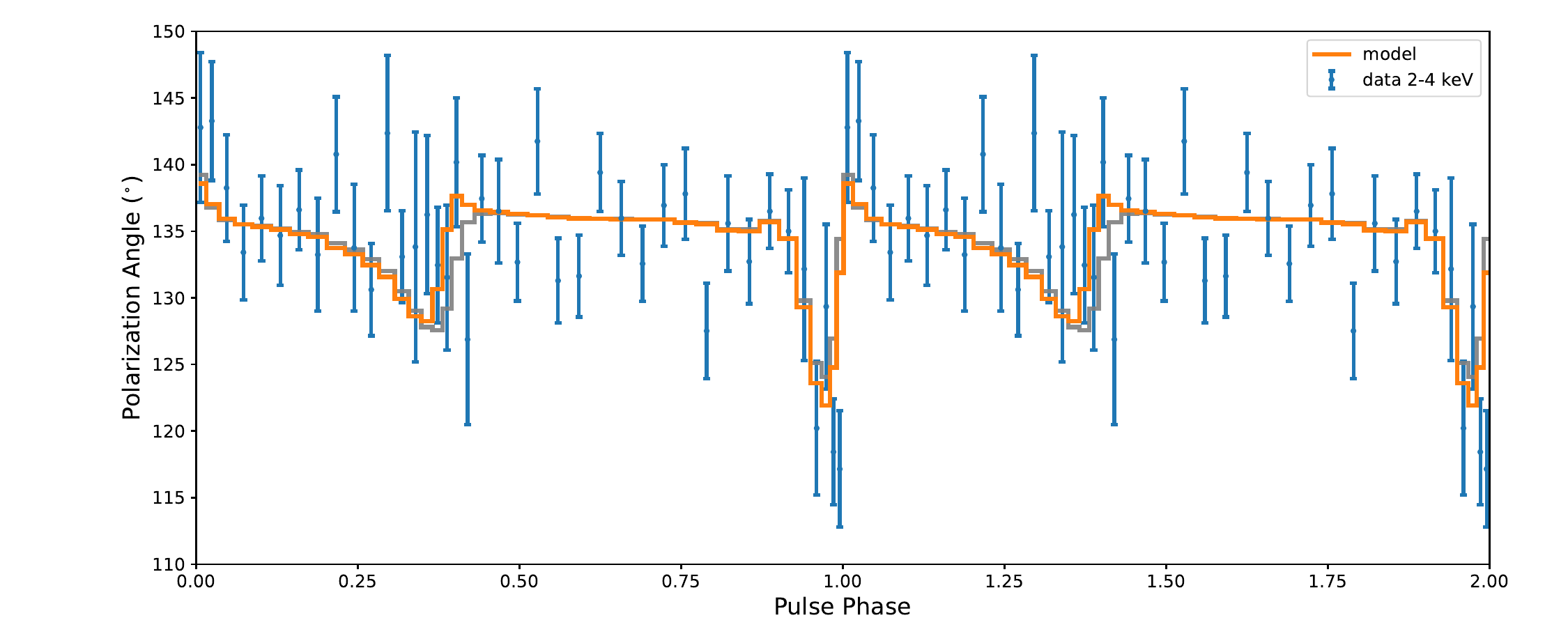}
    \includegraphics[width=0.52\linewidth]{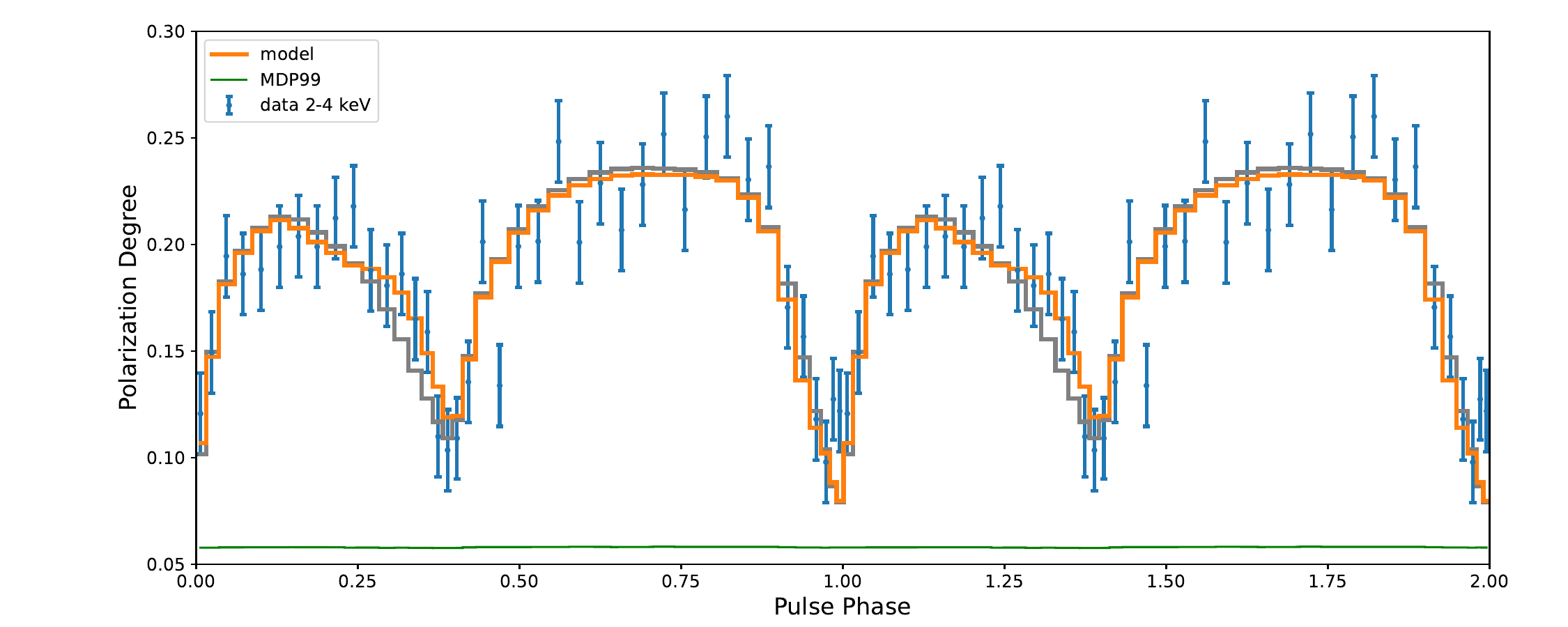}
    \includegraphics[width=0.52\linewidth]{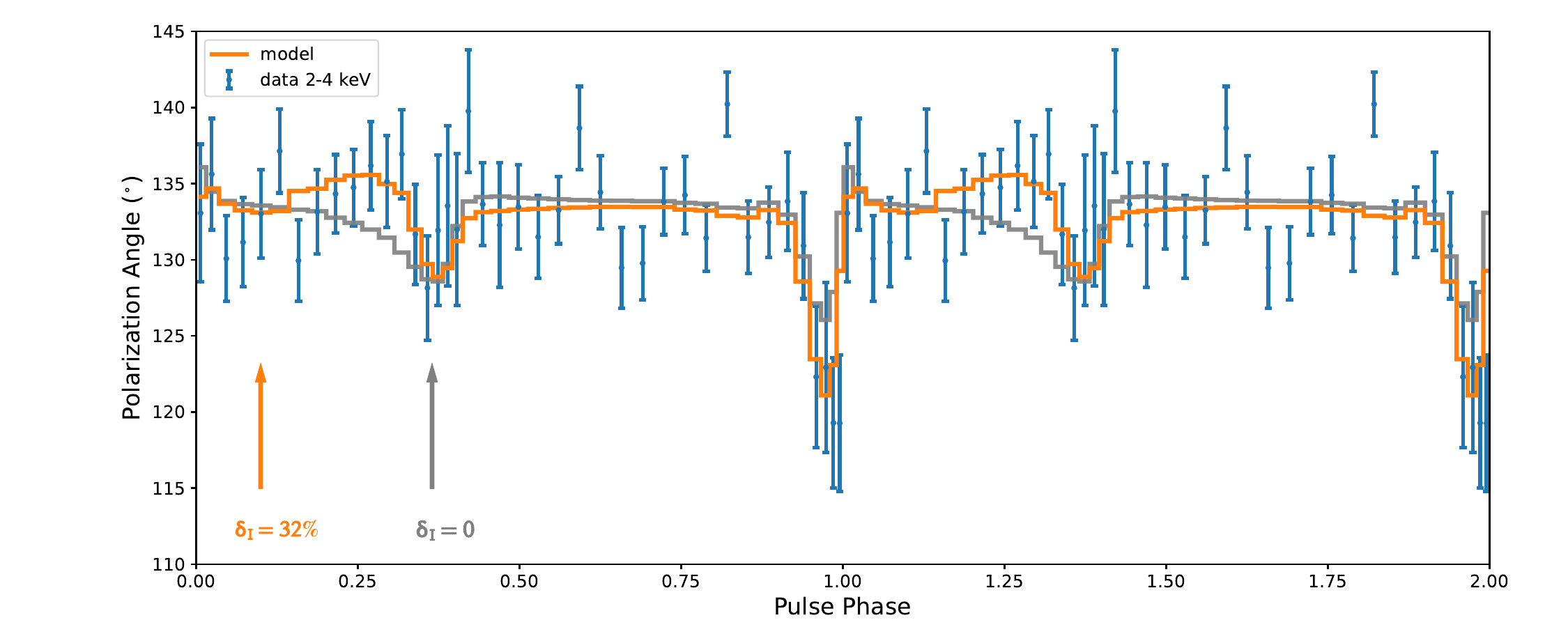}
    \includegraphics[width=0.52\linewidth]{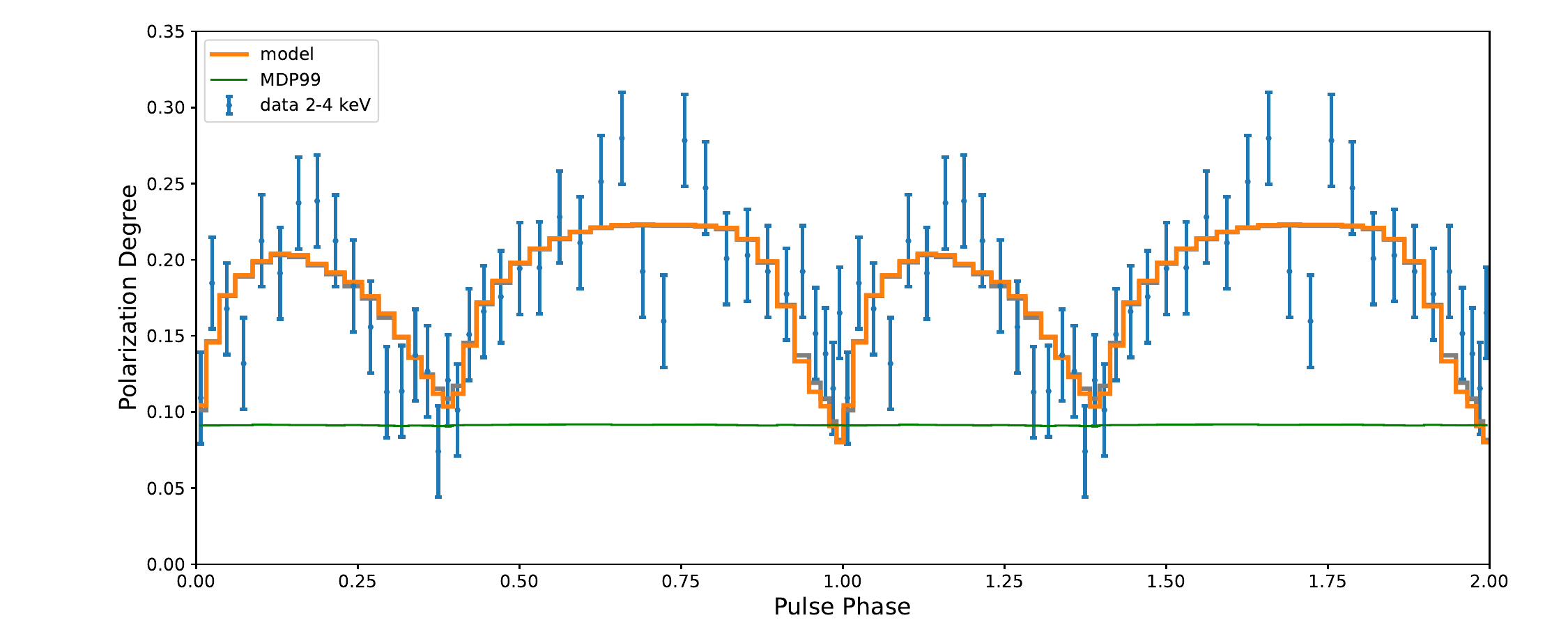}
    \includegraphics[width=0.52\linewidth]{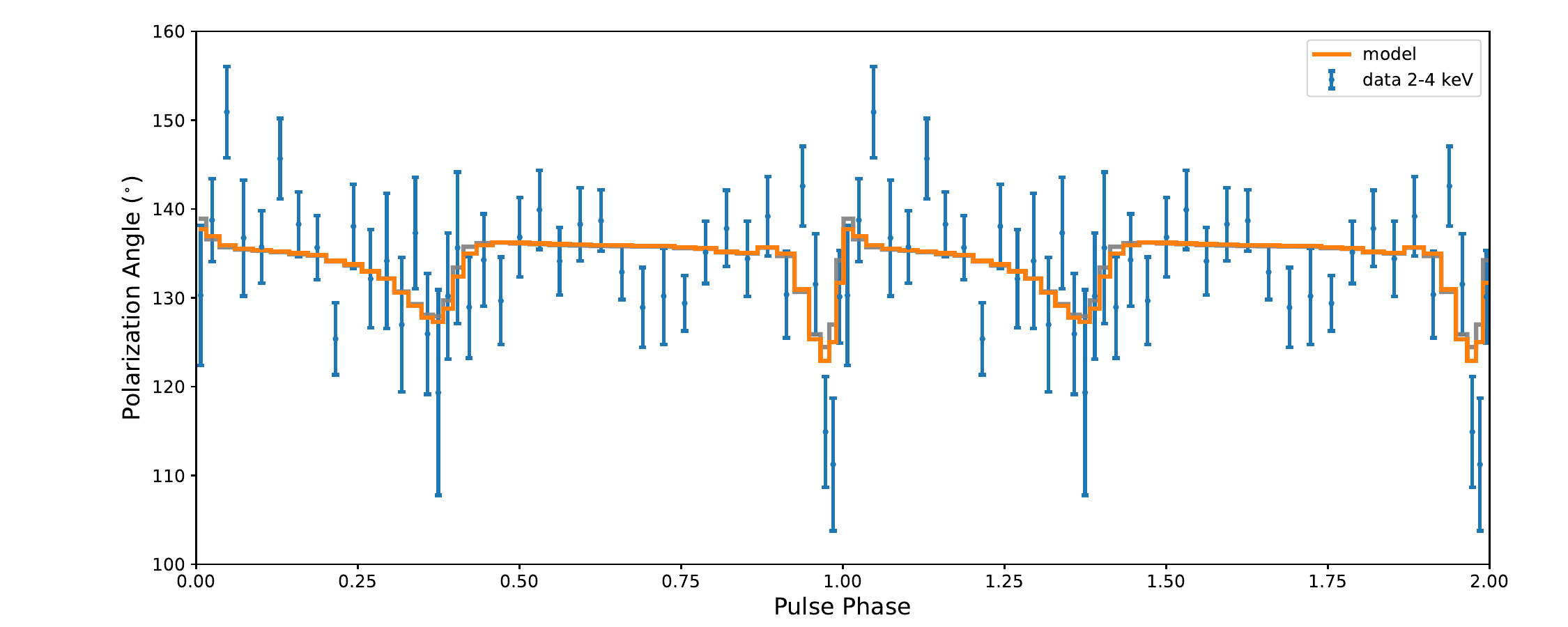}
    \caption{All three IXPE observation of Crab pulsar and the best fitted polarimetric model (subtracting $90\%$ of DC component to the Stokes parameters in the optical band). The first (second), third (fourth) and fifth (sixth) panels correspond to the polarization degree (polarization angle) for the first, second, and third observation of crab pulsar, respectively.  The green line  corresponds to $\mathrm{MDP}_{99}$. The gray solid lines  correspond to the best-fitted model without phase shift. The orange and gray arrows in the fourth panel highlight the polarization angle swing, at the interpulse,  with and without a phase shift, respectively. 
    }
    \label{fig:binned_DC}
\end{figure*}

\begin{figure*}
    \centering
    \includegraphics[width=0.8\linewidth]{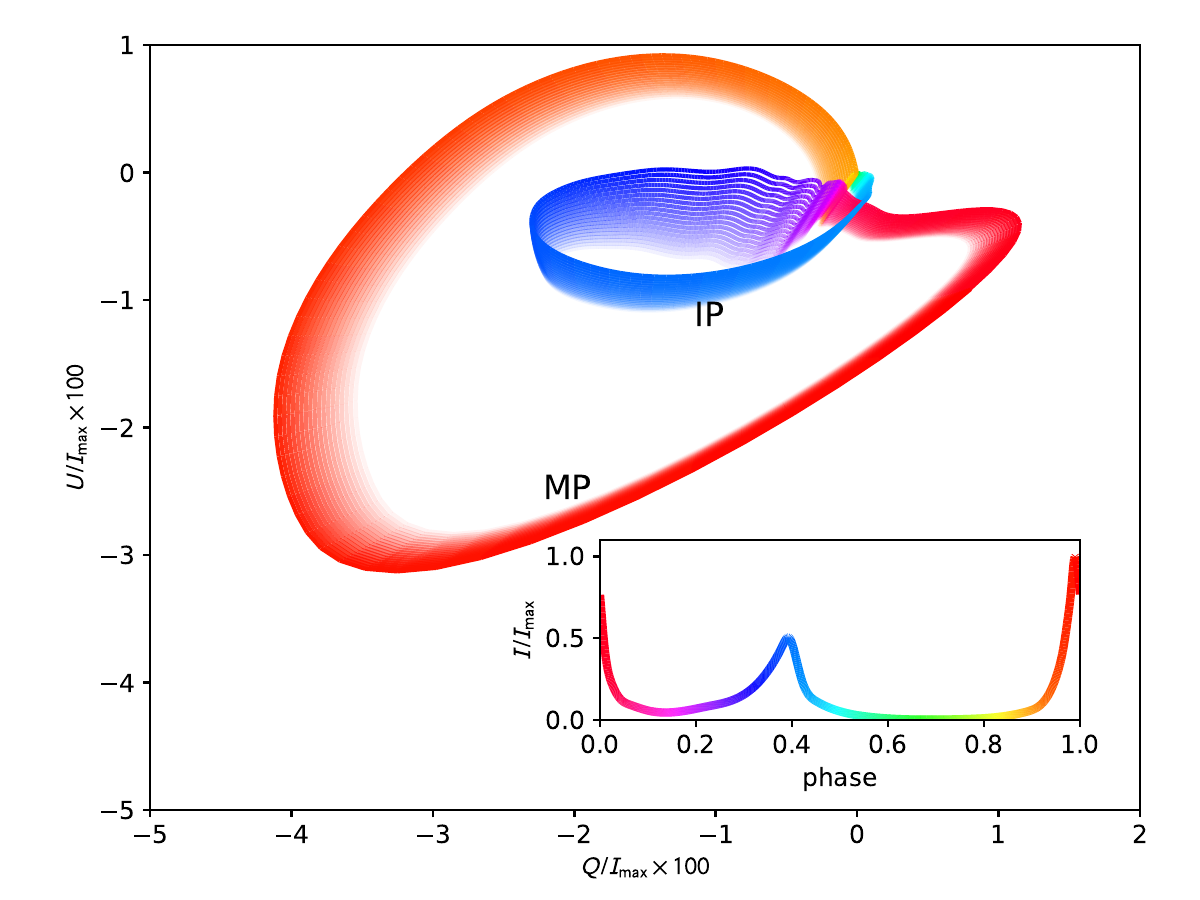}
    \caption{Stokes parameters $Q$ and $U$ for the pulsar alone, extracted from the phenomenological model fitted to IXPE data, without phase-shifts in the polarization angle. The colour gradient corresponds to the pulse phase, as shown in the subplot for the pulse profile (observed by Chandra). The transparency gradient correspond to different DC subtraction from the optical Stokes, ranging from $0\%$ subtraction for the most transparent curve to $90\%$ subtraction for least transparent curve (correspondingly $\beta/\alpha$ ranges from 0.46 to 0.56; see Section 3 for more details).}
    \label{fig:loop}
\end{figure*}

\end{document}